\documentclass[fleqn,usenatbib]{mnras}

\usepackage{newtxtext,newtxmath}

\usepackage[T1]{fontenc}

\DeclareRobustCommand{\VAN}[3]{#2}
\let\VANthebibliography\thebibliography
\def\thebibliography{\DeclareRobustCommand{\VAN}[3]{##3}\VANthebibliography}

\usepackage{graphicx}	
\usepackage{amsmath}	
\usepackage{gensymb}
\usepackage{siunitx}
\usepackage{pdflscape}

\title[Photometric follow-up of eclipsing WD+dM binaries]{Photometric follow-up of 43 new eclipsing white dwarf plus main-sequence binaries from the ZTF survey}

\author[A.~J. Brown et~al.]{
Alex J. Brown,$^{1}$\thanks{E-mail: ajbrown2@sheffield.ac.uk (AJB)}
Steven G. Parsons,$^{1}$
Jan van Roestel,$^{2}$
Alberto Rebassa-Mansergas,$^{3,4}$
Elm\'{e} Breedt,$^{5}$
\newauthor
Vik~S. Dhillon,$^{1,6}$
Martin~J. Dyer,$^{1}$
Matthew~J. Green,$^{7}$
Paul Kerry,$^{1}$
Stuart P. Littlefair$^{1}$
\newauthor
Thomas R.\ Marsh,$^{8}$
James Munday,$^{8,9}$
Ingrid Pelisoli,$^{8}$
David I. Sahman,$^{1}$
James F. Wild,$^{1}$
\\
$^{1}$Department of Physics and Astronomy, Hicks Building, The University of Sheffield, Sheffield, S3 7RH, UK\\
$^{2}$Anton Pannekoek Institute for Astronomy, University of Amsterdam, 1090 GE Amsterdam, The Netherlands\\
$^{3}$Departament de F\'{\i}sica, Universitat Polit\`{e}cnica de Catalunya, c/Esteve Terrades 5, 08860 Castelldefels, Spain\\
$^{4}$Institut d'Estudis Espacials de Catalunya, Ed. Nexus-201, c/Gran Capit\`a 2-4, 08034 Barcelona, Spain\\
$^{5}$Institute of Astronomy, University of Cambridge, Madingley Road, Cambridge CB3 0HA, UK\\
$^{6}$ Instituto de Astrofisica de Canarias, E38205 La Laguna, Tenerife, Spain\\
$^{7}$Department of Astrophysics, School of Physics and Astronomy, Tel Aviv University, Tel Aviv 6997801, Israel\\
$^{8}$Department of Physics, University of Warwick, Gibbet Hill Road, Coventry, CV4 7AL, UK\\
$^{9}$Isaac Newton Group of Telescopes, Apartado de Correos 368, E-38700 Santa Cruz de La Palma, Spain\\
}

\date{Accepted XXX. Received YYY; in original form ZZZ}

\pubyear{2023}

\begin{document}
\label{firstpage}
\pagerange{\pageref{firstpage}--\pageref{lastpage}}
\maketitle

\begin{abstract}
Wide-field time-domain photometric sky surveys are now finding hundreds of eclipsing white dwarf plus M~dwarf binaries, a population encompassing a wealth of information and potential insight into white dwarf and close binary astrophysics.
Precise follow-up observations are essential in order to fully constrain these systems and capitalise on the power of this sample.
We present the first results from our program of high-speed, multi-band photometric follow-up. We develop a method to measure temperatures, (model-dependent) masses, and radii for both components from the eclipse photometry alone and characterize 34 white dwarf binaries, finding general agreement with independent estimates using an alternative approach while achieving around a factor of two increase in parameter precision.
In addition to these parameter estimates, we discover a number of interesting systems -- finding four with sub-stellar secondaries, doubling the number of eclipsing examples, and at least six where we find the white dwarf to be strongly magnetic, making these the first eclipsing examples of such systems and key to investigating the mechanism of magnetic field generation in white dwarfs.
We also discover the first two pulsating white dwarfs in detached and eclipsing post-common-envelope binaries -- one with a low-mass, likely helium core, and one with a relatively high mass, towards the upper end of the known sample of ZZ Cetis. 
Our results demonstrate the power of eclipse photometry, not only as a method of characterising the population, but as a way of discovering important systems that would have otherwise been missed by spectroscopic follow-up.

\end{abstract}




\section{Introduction}

 A significant fraction of field stars are formed as part of a binary system \citep{Eggleton2008, Raghavan2010}.
 Of these binaries, around 25 per cent are formed with sufficiently small orbital separations such that at some stage in their lives the two stars will interact with each other \citep{willems2004}, transferring material between them and affecting their future evolution.
 For many of these interacting systems, the mass-transfer will lead to a common-envelope phase, initiated by the post-main-sequence evolution of the more massive star (the primary).
 This involves both stars being engulfed by the expanding outer envelope of the more massive star, with the resulting drag forces causing the hot core of the primary star and its main-sequence companion to spiral in to small orbital separations and therefore short orbital periods ranging from hours to a few days.
 The lost orbital energy and angular momentum from the binary is imparted into the envelope, ejecting it \citep{Paczynski1976}, where it may then be ionised and lit up by the hot remnant core, appearing as a planetary nebula for a period of time \citep{Jones2017} before the core cools and stratifies to become a white dwarf (WD).
 As well as being a key tracer of the common-envelope phase, these short period detached post-common-envelope binaries (PCEBs) made up of a WD and a main-sequence star are the progenitors to many of the most interesting and exotic astrophysical objects and phenomena in the Universe, including the cosmologically important type Ia supernovae.

Binaries made up of a WD and a main-sequence star are typically split into two categories, one of which containing the WDs with solar-type companions (WD+FGK) and the other made up of WDs with companions of spectral type M and later (referred to as WD+dM).
These categories reflect the differences in observational properties, with the WD+dM binaries being relatively easy to find due to the two stars often contributing a similar amount of flux at optical wavelengths.
This has allowed a large sample to be extracted from spectroscopic surveys \citep{Rebassa-Mansergas2007, Rebassa-Mansergas2010, Rebassa-Mansergas2012b, Rebassa-Mansergas2016}, making them the most common for many years.
More recently, WD+FGK binaries have been found by using UV excesses to discern systems with WDs that would otherwise be outshone by their companion at optical wavelengths \citep{Parsons2016a, Rebassa-Mansergas2017}.

While both types can be found with short orbital periods \citep{Hernandez2022b} -- and therefore with small orbital separations with a relatively high chance of eclipse -- the advantage of the WD+dM PCEBs is that the WD contributes enough of the total flux such that the eclipses can be detected, enabling them to be found in photometric surveys \citep{Parsons2013b, Parsons2015}.
Eclipsing systems are a gold standard in astrophysics, allowing for incredibly precise measurements of the stellar and binary parameters, with typical precisions at or below the percent level.
The result of this is that eclipsing PCEBs are some of the best laboratories of stellar and binary physics available to us and, as such, have been used to test and study a multitude of effects including, but not limited to: precisely measuring mass-radius relations of WDs \citep{Parsons2017a}, confirming the over-inflation of M~dwarfs relative to theoretical models \citep{Parsons2018a}, distinguishing the transition between helium and carbon-oxygen core compositions in WDs \citep{Parsons2017a}, finding systems with brown dwarf companions \citep{Beuermann2013, Parsons2017b, Casewell2020b, vanRoestel2021}, and identifying unusual systems such as merger products and extremely low metallicity systems \citep{O'Brien2001, Rebassa-Mansergas2019a}.

In the current era of wide-field time-domain photometric sky surveys, such as the Zwicky Transient Facility (ZTF; \citealt{Masci2019, Bellm2019, Graham2019}), the number of known eclipsing PCEBs is increasing drastically, with ZTF alone contributing to more than an order of magnitude increase (van Roestel et al. in prep), so far, on the previously known sample \citep{Parsons2015}.
The Legacy Survey of Space and Time (LSST; \citealt{Ivezic2019}) carried out by the Vera Rubin Observatory in the near future will only accelerate this increase.
Follow-up of this vast quantity of systems will be an ongoing challenge, particularly as many of these will be extremely faint, but they will provide much-needed insight into the relatively uncertain physics of the common-envelope as well as uncovering rare systems that may have implications for specific areas of stellar or binary physics.
These include, but are not limited to, systems containing magnetic, pulsating, or high-mass WDs, as well as those with brown dwarf companions.

Previous work has shown that high-cadence multi-colour photometric observations of the primary eclipse is enough to accurately and efficiently characterize detached eclipsing PCEBs \citep{Brown2022}.
This method makes use of the eclipse to cleanly disentangle the spectral energy distributions of the two components and constrains the effective temperatures, while using the shape of the eclipse to measure the orbital inclination and the stellar radii.
These, in turn, provide information about the stellar masses through the use of mass-radius relations.
A photometric method such as this is especially important as fainter systems are discovered -- particularly in the LSST era -- making spectroscopic follow-up even more difficult, and in many cases, impractical.
With this in mind, we have undertaken a program of high-cadence photometric follow-up of eclipsing WD+dM PCEBs (first discovered by van Roestel et al. (in prep)) with the goal of characterizing a significant fraction and discovering a number of rare systems among them. 
Here we present the first results of this follow-up.

\section{Observations}
\subsection{Target selection}
Our targets for follow-up were selected from the detached eclipsing WD+dM systems discovered by van Roestel et al. (in prep) using data from ZTF.
In brief, this sample was created by searching for periodic outliers in the ZTF photometry, indicative of eclipses.
The primary biases are therefore related to the probability of a given system eclipsing as viewed from Earth and the ability to detect an eclipse within the ZTF data.
The former is dominated by the orbital period (with a very weak dependence on the secondary radius), while the latter is dominated by the signal-to-noise ratio of the eclipse, with a heavy dependence on the depth of the eclipse (and a much weaker dependence on the duration of the eclipse).
A more detailed description of the full ZTF eclipsing WD+dM sample identification method and the biases within it will be presented in van Roestel et al. (in prep).

We restricted our target list to systems visible from the La Silla Observatory (Dec < +25~deg) and brighter than $g=19.5~\rm{mag}$.
We typically observed systems with eclipse timings that made for the most efficient use of telescope time on a particular night however we also tried to prioritise systems with longer periods where possible since the eclipses of these systems are more difficult to observe.
Systems with ZTF light curves that indicated they may be of particular interest were also prioritised.
This includes systems with in-eclipse flux measurements at or below the detection threshold of ZTF (indicative of brown dwarf companions) and systems with unusual ZTF light curves, showing variability inconsistent with typical binary variability mechanisms and indicating the presence of a magnetic WD.

A journal of observations is included in \autoref{tab:observations}.

\subsection{High speed photometry}
Our photometric follow-up observations made use of the three-band frame-transfer camera, ULTRACAM \citep{Dhillon2007}, mounted on the 3.6\,m New Technology Telescope (NTT) at the ESO La Silla Observatory in Chile, to obtain high-cadence multi-colour photometry of the primary eclipse of each system -- the eclipse of the WD by its companion.
For all targets observed with ULTRACAM we used the higher throughput Super-SDSS $u_{s}\,g_{s}\,i_{s}$ filters \citep{Dhillon2021}, with the exception of one observation where the $r_{s}$ filter was used in place of $i_{s}$.
For a few of the systems thought to harbour magnetic WDs, we obtained high-speed photometry with the quintuple band frame-transfer camera, HiPERCAM \citep{Dhillon2021}, mounted on the 10.4\,m Gran Telescopio Canar\'ias (GTC) at the Roque de los Muchachos observatory in La Palma, again equipped with Super-SDSS $u_{s}\,g_{s}\,r_{s}\,i_{s}\,z_{s}$ filters.

All observations were bias-subtracted and flat-field corrected (and fringe corrected in the case of the HiPERCAM $z_{s}$ band) using the HiPERCAM pipeline\footnote{\url{https://github.com/HiPERCAM/hipercam}}.
Differential aperture photometry was then extracted using a variable aperture radius set to scale with the measured full width at half-maximum (FWHM) in each frame in order to remove effects due to seeing and transparency variations.
For this we use a target aperture radius of $1.8\times FWHM$. 
In observations with lower signal-to-noise ratios, optimal extraction \citep{Naylor1998} was also performed, with the extraction method resulting in the highest signal-to-noise light curve being the one that was used.

Flux calibration was then performed by fitting the atmospheric extinction in each band using one or more observing runs taken on the same night as the target observations (each spanning a minimum of 0.2 airmasses).
The atmospheric extinction measurements were combined with an observation of an ULTRACAM flux standard star \citep[see][table A3]{Brown2022}, reduced using a larger target aperture radius of $2.5\times FWHM$, in order to measure the instrumental zeropoint for the night.
The calibrated flux of the comparison star was then determined using the same target aperture radius as for the flux standard star, which was then used to flux calibrate the target.
When using optimally extracted photometry, the flux calibration was still performed on the data reduced using a standard aperture photometry extraction.
This calibration was then applied to the optimally extracted photometry to prevent systematic absolute flux errors between the two methods.
These flux calibration steps were performed using the \textsc{cam\_cal}\footnote{\url{https://github.com/Alex-J-Brown/cam_cal}} package.

\section{Method}
We fit the flux calibrated eclipse photometry using the \textsc{pylcurve}\footnote{\url{https://github.com/Alex-J-Brown/pylcurve}} package, a python wrapper for \textsc{lcurve's} \textsc{lroche} routine \citep{Copperwheat2010}.
In general, we follow the method of \citet{Brown2022} which involves fitting the eclipse photometry in multiple filters simultaneously with eight free parameters.
These are the effective temperatures, $\rm{T_{1}}$ and $\rm{T_{2}}$, which define the spectral energy distributions (SEDs) of both stars through the use of stellar atmosphere models \citep{Claret2020a, Husser2013}; the stellar masses, $\rm{M_{1}}$ and $\rm{M_{2}}$; the binary inclination, $i$; the parallax, $\varpi$; the interstellar reddening, $E(B-V)$; and the time of mid-eclipse, $\rm{T_{0}}$.
With the use of mass-radius relations and a given (fixed) orbital period, $\rm{P}$, the radii of both stars and the orbital separation of the binary can be defined allowing model light curves to be generated for each filter.
See \citet{Brown2022} for more details on this method.

For this work, however, we implement two changes to the methodology mentioned above, both regarding the spectral modelling of the secondary star: 
\begin{enumerate}
    \item \label{itm:first} Previously, PHOENIX stellar atmospheres \citep{Husser2013} were used to model the SED of the secondary star \citep{Brown2022}.
    However, these models are limited to a minimum effective temperature of 2300~K, preventing the modelling of systems with brown dwarf companions.
    We have therefore switched to using the BT-Settl CIFIST stellar atmosphere grid \citep{Allard2012} which go as low as 1200~K, allowing for a seamless transition to the brown dwarf regime and keeping our modelling consistent throughout.
    
    \item \label{itm:second} It is well known that there are significant differences in the synthetic photometry of low mass stars calculated using different spectral models for a given effective temperature and surface gravity.
    This is most apparent for lower effective temperatures (<3500~K), with models struggling to reproduce the transitions from M~dwarfs to L~dwarfs to T~dwarfs \citep{Saumon2008, Allard2012, Best2021}.
    Rigidly defining the SED of the secondary from these spectral models could therefore introduce problems where the model photometry cannot reproduce the observed SED of the star in question to the precision of our observations.
    We counter this by allowing the secondary to have a separate effective temperature in each observed bandpass.
    Despite being allowed to vary, these individual filter-specific effective temperatures should be consistent with each other at a certain level.
    We implement this consistency requirement using priors to favour solutions where these effective temperatures are similar across the different filters.
\end{enumerate} 

\begin{figure}
 \includegraphics[width=\columnwidth]{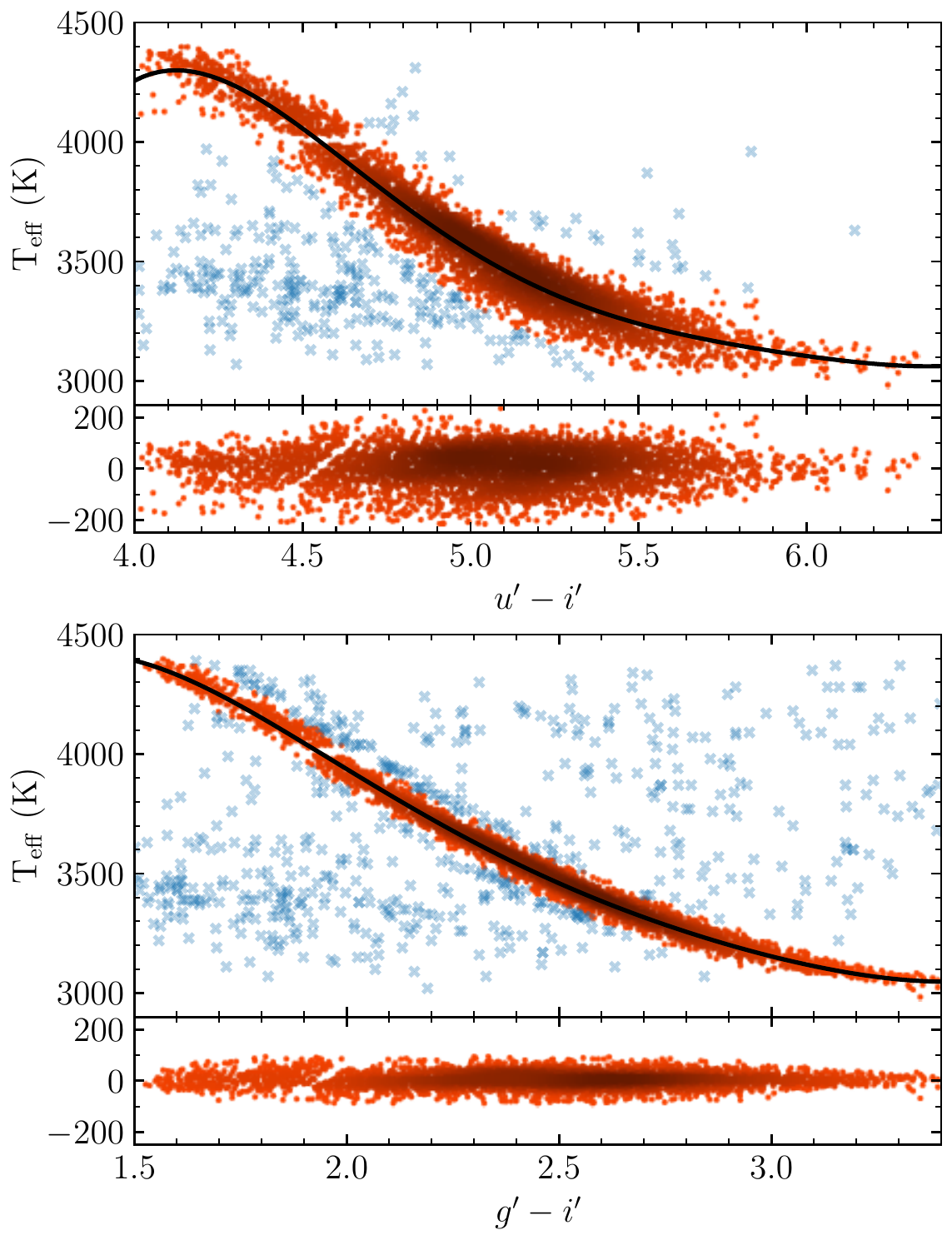}
 \caption{Effective temperatures of M~dwarfs measured by \citet{Morrell2019} against their SDSS colours. Blue crosses show points discarded by the sigma clipping procedure and the solid black lines show the final polynomial fits to these sigma-clipped distributions. The residuals of these fits, from which we calculate the standard deviations, are shown in the panels below. The gap in the sample at an effective temperature of $4000~\rm{K}$ is due to a discontinuity in the model grid used by \citet{Morrell2019}.}
 \label{fig:temp_var}
\end{figure}

In order to inform the priors on the filter-specific secondary temperatures mentioned in \autoref{itm:second}, we use a sample of 15\,279 well-characterised M~dwarfs \citep{Morrell2019}.
Cross-matching this sample with SDSS DR13 returns a sample of 5\,222 M~dwarfs, on which we then make colour cuts informed by synthetic photometry of the BT-SETTL-CIFIST model atmospheres ($4.0 < (u'-i') < 6.4$ and $1.5 < (g'-i') < 3.4$) to remove many of the extreme outliers. This leaves 4\,158 M~dwarfs with SDSS photometry.
We then fit fifth-order polynomials to the measured effective temperature as a function of $u'-i'$ and $g'-i'$ colours individually, using an iterative sigma clipping procedure with a $3\sigma$ cut to remove any outliers that remain after the initial colour cuts (\autoref{fig:temp_var}).
The standard deviations of the residuals of the remaining points are 80~K for a $u'-i'$ colour and 30~K for a $g'-i'$ colour.
We therefore implement Gaussian priors on the difference in effective temperature between the $u'$ and $i'$, and $g'$ and $i'$ bands of 80~K and 30~K respectively, both centred at zero.
As, with this method, there are as many temperature measurements available for the secondary as filters used, we take the $i_{s}$-band measurement as being representative of the true secondary temperature.
We make this choice based on it being the the band where the secondary is brightest and is therefore the most strongly constrained by the photometry.

As in \citet{Brown2022}, we use a Markov Chain Monte Carlo (MCMC) method to fit each light curve, implemented through the \textsc{python} package, \textsc{emcee}\footnote{\url{https://emcee.readthedocs.io/en/stable/}} \citep{Foreman-Mackey2013}.
We run each fit for a minimum of 10\,000 steps using 100 walkers and inspect each fit manually for convergence and stability.
Each system is first fit using a carbon-oxygen (CO) core WD mass-radius relation \citep{Bedard2020, Blouin2018, Tremblay2011} with the fit then being repeated using a helium (He) core model \citep{Panei2007} if the best-fit CO-core WD mass is below $0.5~\rm{M_{\odot}}$.
If this subsequent fit using the He-core model is restricted by the upper mass limit of the He-core models -- $0.5~\rm{M_{\odot}}$ -- then we consider the WD to have a CO core-composition, if not then we assume the WD to possess a He core.

\section{Results}

The results of our light curve fits are presented in \autoref{tab:stellar_results} and \autoref{tab:binary_results} -- note that of the 43 systems that we have followed-up, 9 do not have measured parameters because they either harbour magnetic WDs or are strong candidates (see section \ref{sec:magnetic_wd}).
Our best-fit values are taken to be the median of the posterior distributions of the MCMC with lower and upper uncertainties taken as the 16th and 84th percentiles respectively.
As in \citet{Brown2022}, the formal uncertainties from the MCMC do not include contributions from systematic errors and so we attempt to take this into account by adding estimated systematic uncertainties in quadrature with the formal uncertainties of the MCMC.
We add 1.5~per~cent in quadrature with the uncertainties on the primary temperature \citep{Gianninas2011}, $\rm{T_{1}}$, and 100~K in quadrature with the secondary temperature, $\rm{T_{2}}$. We also add 1~per~cent in quadrature with the WD mass, $\rm{M_{1}}$, and 5 per cent in quadrature with the secondary mass, $\rm{M_{2}}$ (for the reasons explained in \citet{Brown2022}).
These contributions are included in the uncertainties shown in \autoref{tab:stellar_results} and in all figures. 
An example ULTRACAM eclipse light curve and best-fit model is shown in \autoref{fig:lc_example} with all best-fit light curves shown in Appendix \ref{sec:light_curves}.

\begin{figure}
 \includegraphics[width=\columnwidth]{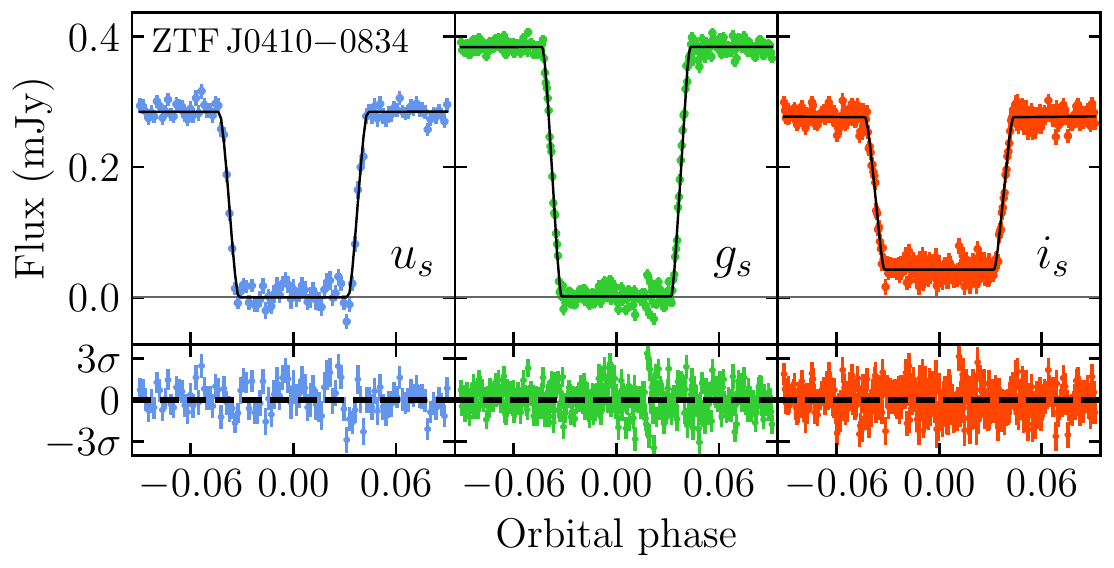}
 \caption{ULTRACAM $u_{s}\,g_{s}\,i_{s}$ eclipse light curve (coloured points) of ZTF\,J041016.82$-$083419.5 with the best-fit light curve model over-plotted in black and the residuals of this fit shown below. The zero-flux level is shown by the horizontal grey line.}
 \label{fig:lc_example}
\end{figure}

\begin{table*}
\centering
\caption{Best fit stellar parameters to the ULTRACAM eclipse photometry. Uncertainties include estimated systematic errors added in quadrature with the formal uncertainties of the MCMC. These estimated systematics are 1.5~per~cent on $\rm{T_{1}}$ \citep{Gianninas2011}, 100~K on $\rm{T_{2}}$, 1 per cent for $\rm{M_{1}}$, and 5 per cent for $\rm{M_{2}}$ \citep{Brown2022}.}
\tabcolsep=0.13cm
\label{tab:stellar_results}
\begin{tabular}{@{}lcllllccll@{}}
\hline
Target & He/CO & $\rm{T_{1}~(K)}$ & $\rm{M_{1}~(M_{\odot})}$ & $\rm{R_{1}~(R_{\odot})}$ & $\rm{log(g_{1})}$ & $\rm{T_{2}~(K)}$ & $\rm{M_{2}~(M_{\odot})}$ & $\rm{R_{2}~(R_{\odot})}$ & $\rm{R_{2}/R_{L1}}$ \\
\hline
ZTF\,J041016.82$-$083419.5 & He & $14690^{+560}_{-550}$ & $0.355^{+0.015}_{-0.011}$ & $0.0204^{+0.0004}_{-0.0006}$ & $7.37^{+0.04}_{-0.03}$ & $2840^{+110}_{-110}$ & $0.123^{+0.009}_{-0.008}$ & $0.151^{+0.008}_{-0.006}$ & $0.680^{+0.033}_{-0.021}$ \\[0.8ex]
ZTF\,J051902.06$+$092526.4 & He & $10750^{+770}_{-580}$ & $0.391^{+0.019}_{-0.029}$ & $0.0178^{+0.0009}_{-0.0005}$ & $7.53^{+0.04}_{-0.08}$ & $2800^{+140}_{-110}$ & $0.177^{+0.014}_{-0.019}$ & $0.214^{+0.012}_{-0.018}$ & $0.842^{+0.079}_{-0.084}$ \\[0.8ex]
ZTF\,J052848.24$+$215629.0 & CO & $12100^{+700}_{-630}$ & $0.787^{+0.025}_{-0.025}$ & $0.0105^{+0.0003}_{-0.0003}$ & $8.29^{+0.04}_{-0.04}$ & $3130^{+110}_{-110}$ & $0.184^{+0.014}_{-0.013}$ & $0.220^{+0.011}_{-0.009}$ & $0.408^{+0.014}_{-0.011}$ \\[0.8ex]
ZTF\,J053708.26$-$245014.6 & He & $16100^{+440}_{-410}$ & $0.397^{+0.009}_{-0.007}$ & $0.0191^{+0.0002}_{-0.0002}$ & $7.48^{+0.02}_{-0.02}$ & $2970^{+100}_{-100}$ & $0.204^{+0.012}_{-0.011}$ & $0.241^{+0.007}_{-0.005}$ & $0.333^{+0.006}_{-0.004}$ \\[0.8ex]
ZTF\,J061530.96$+$051041.8 & CO & $15220^{+600}_{-510}$ & $0.560^{+0.011}_{-0.011}$ & $0.0139^{+0.0002}_{-0.0002}$ & $7.90^{+0.02}_{-0.02}$ & $3380^{+110}_{-110}$ & $0.533^{+0.030}_{-0.029}$ & $0.547^{+0.013}_{-0.011}$ & $0.531^{+0.008}_{-0.008}$ \\[0.8ex]
ZTF\,J063808.71$+$091027.4 & CO & $22500^{+1200}_{-1000}$ & $0.604^{+0.013}_{-0.011}$ & $0.0136^{+0.0001}_{-0.0002}$ & $7.95^{+0.02}_{-0.02}$ & $3320^{+110}_{-110}$ & $0.410^{+0.024}_{-0.022}$ & $0.432^{+0.012}_{-0.008}$ & $0.295^{+0.005}_{-0.004}$ \\[0.8ex]
ZTF\,J063954.70$+$191958.0 & CO & $15980^{+520}_{-520}$ & $0.701^{+0.011}_{-0.009}$ & $0.0117^{+0.0001}_{-0.0001}$ & $8.15^{+0.01}_{-0.01}$ & $3200^{+100}_{-100}$ & $0.210^{+0.011}_{-0.011}$ & $0.246^{+0.004}_{-0.002}$ & $0.398^{+0.004}_{-0.002}$ \\[0.8ex]
ZTF\,J064242.41$+$131427.6 & CO & $14560^{+540}_{-500}$ & $0.633^{+0.011}_{-0.008}$ & $0.0127^{+0.0001}_{-0.0001}$ & $8.03^{+0.02}_{-0.01}$ & $3110^{+100}_{-100}$ & $0.150^{+0.008}_{-0.008}$ & $0.183^{+0.004}_{-0.001}$ & $0.438^{+0.006}_{-0.002}$ \\[0.8ex]
ZTF\,J065103.70$+$145246.2 & CO & $13140^{+560}_{-670}$ & $0.515^{+0.019}_{-0.020}$ & $0.0145^{+0.0003}_{-0.0003}$ & $7.83^{+0.03}_{-0.04}$ & $3170^{+120}_{-110}$ & $0.242^{+0.018}_{-0.019}$ & $0.276^{+0.012}_{-0.013}$ & $0.589^{+0.018}_{-0.019}$ \\[0.8ex]
ZTF\,J070458.08$-$020103.3 & CO & $9280^{+230}_{-250}$ & $0.500^{+0.012}_{-0.015}$ & $0.0143^{+0.0003}_{-0.0002}$ & $7.82^{+0.02}_{-0.03}$ & $3300^{+100}_{-100}$ & $0.344^{+0.018}_{-0.020}$ & $0.370^{+0.006}_{-0.010}$ & $0.915^{+0.040}_{-0.043}$ \\[0.8ex]
ZTF\,J071759.04$+$113630.2 & CO & $21110^{+920}_{-750}$ & $0.528^{+0.016}_{-0.017}$ & $0.0149^{+0.0003}_{-0.0003}$ & $7.81^{+0.03}_{-0.03}$ & $3150^{+120}_{-110}$ & $0.296^{+0.020}_{-0.022}$ & $0.326^{+0.013}_{-0.015}$ & $0.320^{+0.008}_{-0.009}$ \\[0.8ex]
ZTF\,J071843.68$-$085232.1 & CO & $18940^{+870}_{-880}$ & $0.794^{+0.019}_{-0.018}$ & $0.0106^{+0.0002}_{-0.0002}$ & $8.28^{+0.03}_{-0.03}$ & $3120^{+110}_{-110}$ & $0.306^{+0.020}_{-0.019}$ & $0.335^{+0.012}_{-0.011}$ & $0.555^{+0.014}_{-0.012}$ \\[0.8ex]
ZTF\,J080441.95$-$021545.7 & CO & $13430^{+560}_{-550}$ & $0.577^{+0.010}_{-0.009}$ & $0.0134^{+0.0001}_{-0.0001}$ & $7.94^{+0.01}_{-0.01}$ & $<1510^{+260}_{-200}$ & $<0.069^{+0.007}_{-0.007}$ & $0.098^{+0.002}_{-0.001}$ & $0.377^{+0.008}_{-0.006}$ \\[0.8ex]
ZTF\,J080542.98$-$143036.3 & He & $26500^{+1200}_{-9000}$ & $0.393^{+0.013}_{-0.013}$ & $0.0239^{+0.0007}_{-0.0006}$ & $7.28^{+0.03}_{-0.04}$ & $3250^{+120}_{-110}$ & $0.291^{+0.020}_{-0.023}$ & $0.331^{+0.013}_{-0.017}$ & $0.586^{+0.016}_{-0.021}$ \\[0.8ex]
ZTF\,J094826.35$+$253810.6 & CO & $11290^{+480}_{-450}$ & $0.504^{+0.026}_{-0.024}$ & $0.0145^{+0.0004}_{-0.0004}$ & $7.82^{+0.05}_{-0.05}$ & $3120^{+120}_{-120}$ & $0.169^{+0.015}_{-0.014}$ & $0.205^{+0.013}_{-0.012}$ & $0.546^{+0.024}_{-0.024}$ \\[0.8ex]
ZTF\,J102254.00$-$080327.3 & CO & $8330^{+260}_{-250}$ & $0.605^{+0.027}_{-0.025}$ & $0.0127^{+0.0003}_{-0.0003}$ & $8.01^{+0.04}_{-0.04}$ & $3170^{+110}_{-110}$ & $0.405^{+0.030}_{-0.029}$ & $0.428^{+0.021}_{-0.020}$ & $0.620^{+0.023}_{-0.021}$ \\[0.8ex]
ZTF\,J102653.47$-$101330.3 & He & $19320^{+710}_{-670}$ & $0.376^{+0.012}_{-0.010}$ & $0.0214^{+0.0004}_{-0.0007}$ & $7.35^{+0.04}_{-0.02}$ & $2840^{+110}_{-110}$ & $0.105^{+0.008}_{-0.006}$ & $0.134^{+0.007}_{-0.004}$ & $0.558^{+0.021}_{-0.012}$ \\[0.8ex]
ZTF\,J103448.82$+$005201.9 & He & $10060^{+410}_{-370}$ & $0.455^{+0.007}_{-0.007}$ & $0.0159^{+0.0001}_{-0.0001}$ & $7.69^{+0.01}_{-0.01}$ & $<1550^{+250}_{-230}$ & $<0.067^{+0.005}_{-0.006}$ & $0.097^{+0.001}_{-0.001}$ & $0.460^{+0.010}_{-0.008}$ \\[0.8ex]
ZTF\,J104906.96$-$175530.7 & He & $13000^{+440}_{-460}$ & $0.426^{+0.010}_{-0.007}$ & $0.0173^{+0.0001}_{-0.0002}$ & $7.59^{+0.02}_{-0.01}$ & $3170^{+100}_{-110}$ & $0.198^{+0.012}_{-0.010}$ & $0.235^{+0.007}_{-0.003}$ & $0.402^{+0.008}_{-0.003}$ \\[0.8ex]
ZTF\,J122009.98$+$082155.0 & CO & $10170^{+270}_{-260}$ & $0.580^{+0.017}_{-0.018}$ & $0.0132^{+0.0003}_{-0.0002}$ & $7.96^{+0.03}_{-0.03}$ & $3140^{+110}_{-110}$ & $0.275^{+0.019}_{-0.020}$ & $0.306^{+0.012}_{-0.013}$ & $0.157^{+0.004}_{-0.004}$ \\[0.8ex]
ZTF\,J125620.57$+$211725.8 & CO & $5073^{+79}_{-79}$ & $0.479^{+0.010}_{-0.009}$ & $0.0141^{+0.0001}_{-0.0001}$ & $7.82^{+0.02}_{-0.01}$ & $2950^{+100}_{-100}$ & $0.101^{+0.005}_{-0.005}$ & $0.125^{+0.001}_{-0.001}$ & $0.152^{+0.001}_{-0.001}$ \\[0.8ex]
ZTF\,J130228.34$-$003200.2 & CO & $11790^{+400}_{-330}$ & $0.811^{+0.021}_{-0.016}$ & $0.0102^{+0.0002}_{-0.0002}$ & $8.33^{+0.03}_{-0.02}$ & $3030^{+100}_{-100}$ & $0.179^{+0.012}_{-0.010}$ & $0.216^{+0.008}_{-0.005}$ & $0.502^{+0.013}_{-0.009}$ \\[0.8ex]
ZTF\,J134151.70$-$062613.9 & CO & $58300^{+8400}_{-8700}$ & $0.509^{+0.038}_{-0.035}$ & $0.0225^{+0.0009}_{-0.0016}$ & $7.43^{+0.09}_{-0.04}$ & $2800^{+210}_{-220}$ & $0.126^{+0.015}_{-0.009}$ & $0.159^{+0.018}_{-0.007}$ & $0.617^{+0.062}_{-0.021}$ \\[0.8ex]
ZTF\,J140036.65$+$081447.4 & CO & $13340^{+650}_{-610}$ & $0.563^{+0.009}_{-0.008}$ & $0.0137^{+0.0001}_{-0.0001}$ & $7.92^{+0.01}_{-0.01}$ & $2970^{+100}_{-100}$ & $0.232^{+0.012}_{-0.012}$ & $0.268^{+0.003}_{-0.001}$ & $0.418^{+0.003}_{-0.001}$ \\[0.8ex]
ZTF\,J140423.86$+$065557.7 & CO & $14980^{+470}_{-460}$ & $0.736^{+0.016}_{-0.015}$ & $0.0113^{+0.0002}_{-0.0002}$ & $8.20^{+0.02}_{-0.02}$ & $3100^{+100}_{-100}$ & $0.409^{+0.023}_{-0.023}$ & $0.432^{+0.010}_{-0.010}$ & $0.884^{+0.045}_{-0.031}$ \\[0.8ex]
ZTF\,J140537.34$+$103919.0 & He & $29900^{+9000}_{-1100}$ & $0.404^{+0.008}_{-0.008}$ & $0.0279^{+0.0006}_{-0.0006}$ & $7.15^{+0.02}_{-0.02}$ & $3430^{+130}_{-140}$ & $0.085^{+0.005}_{-0.005}$ & $0.112^{+0.003}_{-0.003}$ & $0.234^{+0.004}_{-0.004}$ \\[0.8ex]
ZTF\,J140702.57$+$211559.7 & He & $10870^{+350}_{-350}$ & $0.406^{+0.018}_{-0.014}$ & $0.0173^{+0.0004}_{-0.0004}$ & $7.57^{+0.04}_{-0.03}$ & $3160^{+110}_{-110}$ & $0.263^{+0.021}_{-0.016}$ & $0.296^{+0.015}_{-0.009}$ & $0.702^{+0.029}_{-0.016}$ \\[0.8ex]
ZTF\,J145819.54$+$131326.7 & CO & $9420^{+260}_{-260}$ & $0.581^{+0.010}_{-0.010}$ & $0.0131^{+0.0001}_{-0.0001}$ & $7.97^{+0.01}_{-0.01}$ & $<1730^{+240}_{-270}$ & $<0.067^{+0.006}_{-0.006}$ & $0.095^{+0.001}_{-0.000}$ & $0.446^{+0.011}_{-0.006}$ \\[0.8ex]
ZTF\,J162644.18$-$101854.3 & CO & $36700^{+2700}_{-2700}$ & $0.499^{+0.015}_{-0.012}$ & $0.0180^{+0.0002}_{-0.0003}$ & $7.62^{+0.02}_{-0.01}$ & $3180^{+110}_{-110}$ & $0.212^{+0.013}_{-0.011}$ & $0.259^{+0.008}_{-0.003}$ & $0.425^{+0.008}_{-0.004}$ \\[0.8ex]
ZTF\,J163421.00$-$271321.7 & He & $10680^{+790}_{-630}$ & $0.436^{+0.042}_{-0.054}$ & $0.0166^{+0.0013}_{-0.0009}$ & $7.64^{+0.09}_{-0.12}$ & $2400^{+130}_{-120}$ & $0.134^{+0.016}_{-0.020}$ & $0.163^{+0.019}_{-0.022}$ & $0.759^{+0.128}_{-0.099}$ \\[0.8ex]
ZTF\,J164441.18$+$243428.2 & He & $13270^{+520}_{-460}$ & $0.382^{+0.020}_{-0.018}$ & $0.0188^{+0.0007}_{-0.0007}$ & $7.47^{+0.05}_{-0.05}$ & $2500^{+110}_{-110}$ & $0.103^{+0.009}_{-0.009}$ & $0.129^{+0.009}_{-0.008}$ & $0.607^{+0.033}_{-0.028}$ \\[0.8ex]
ZTF\,J180256.45$-$005458.3 & He & $10770^{+630}_{-500}$ & $0.458^{+0.019}_{-0.021}$ & $0.0160^{+0.0004}_{-0.0003}$ & $7.69^{+0.03}_{-0.04}$ & $3150^{+110}_{-110}$ & $0.150^{+0.010}_{-0.011}$ & $0.182^{+0.008}_{-0.010}$ & $0.319^{+0.010}_{-0.012}$ \\[0.8ex]
ZTF\,J182848.77$+$230838.0 & CO & $16620^{+560}_{-650}$ & $0.594^{+0.009}_{-0.008}$ & $0.0134^{+0.0001}_{-0.0001}$ & $7.96^{+0.01}_{-0.01}$ & $<2290^{+110}_{-120}$ & $<0.068^{+0.007}_{-0.006}$ & $0.096^{+0.002}_{-0.000}$ & $0.392^{+0.009}_{-0.005}$ \\[0.8ex]
ZTF\,J195456.71$+$101937.5 & CO & $21500^{+1000}_{-1100}$ & $0.509^{+0.015}_{-0.012}$ & $0.0154^{+0.0002}_{-0.0002}$ & $7.77^{+0.03}_{-0.02}$ & $3480^{+110}_{-110}$ & $0.449^{+0.028}_{-0.026}$ & $0.470^{+0.016}_{-0.013}$ & $0.523^{+0.012}_{-0.010}$ \\[0.8ex]
\hline
\end{tabular}
\end{table*}

\begin{table*}
\centering
\caption{Best fit binary parameters to the ULTRACAM eclipse photometry. The orbital periods are listed here for reference but are not fitted parameters and so do not have corresponding uncertainties. The \textit{Gaia} DR3 parallax measurements are included for comparison.}
\tabcolsep=0.13cm
\label{tab:binary_results}
\begin{tabular}{@{}llllllll@{}}
\hline
Target & $i(\degree)$ & $\rm{a~(R_{\odot})}$ & $E(B-V)$ & $\rm{\varpi_{UCAM}}$ & $\rm{\varpi_{Gaia}}$ & $\rm{T_{0}~(BMJD(TDB))}$ & $\rm{P~(d)}$ \\
\hline
ZTF\,J041016.82$-$083419.5 & $86.6^{+2.1}_{-1.7}$ & $0.616^{+0.009}_{-0.006}$ & $0.031^{+0.017}_{-0.017}$ & $3.863^{+0.091}_{-0.078}$ & $4.07\pm0.11$ & 59646.0489782(16) & 0.0811093 \\ [0.8ex]
ZTF\,J051902.06$+$092526.4 & $76.3^{+1.1}_{-0.6}$ & $0.715^{+0.012}_{-0.020}$ & $0.112^{+0.028}_{-0.023}$ & $2.835^{+0.140}_{-0.140}$ & $2.92\pm0.30$ & 59251.0519387(57) & 0.0929131 \\ [0.8ex]
ZTF\,J052848.24$+$215629.0 & $87.7^{+1.4}_{-1.0}$ & $1.546^{+0.017}_{-0.016}$ & $0.090^{+0.020}_{-0.021}$ & $5.666^{+0.104}_{-0.111}$ & $5.59\pm0.13$ & 59932.215321(52) & 0.2259952 \\ [0.8ex]
ZTF\,J053708.26$-$245014.6 & $88.1^{+0.7}_{-0.6}$ & $1.688^{+0.014}_{-0.010}$ & $0.015^{+0.011}_{-0.010}$ & $4.580^{+0.044}_{-0.047}$ & $4.574\pm0.049$ & 59251.2246115(52) & 0.3277936 \\ [0.8ex]
ZTF\,J061530.96$+$051041.8 & $85.0^{+0.7}_{-0.7}$ & $2.146^{+0.015}_{-0.014}$ & $0.019^{+0.019}_{-0.013}$ & $3.163^{+0.060}_{-0.051}$ & $3.166\pm0.081$ & 59280.12536567(83) & 0.3481742 \\ [0.8ex]
ZTF\,J063808.71$+$091027.4 & $88.2^{+0.5}_{-0.6}$ & $3.197^{+0.024}_{-0.019}$ & $0.021^{+0.018}_{-0.015}$ & $1.709^{+0.047}_{-0.047}$ & $1.65\pm0.14$ & 59252.1564861(10) & 0.6576453 \\ [0.8ex]
ZTF\,J063954.70$+$191958.0 & $88.9^{+0.7}_{-0.7}$ & $1.659^{+0.008}_{-0.005}$ & $0.028^{+0.013}_{-0.015}$ & $5.394^{+0.070}_{-0.075}$ & $5.387\pm0.085$ & 59251.17799186(52) & 0.2593556 \\ [0.8ex]
ZTF\,J064242.41$+$131427.6 & $89.1^{+0.7}_{-0.8}$ & $1.195^{+0.006}_{-0.003}$ & $0.022^{+0.016}_{-0.013}$ & $3.583^{+0.075}_{-0.073}$ & $3.77\pm0.20$ & 59252.10345653(59) & 0.1710542 \\ [0.8ex]
ZTF\,J065103.70$+$145246.2 & $85.3^{+1.5}_{-1.0}$ & $1.166^{+0.016}_{-0.017}$ & $0.037^{+0.016}_{-0.018}$ & $2.567^{+0.073}_{-0.076}$ & $2.70\pm0.17$ & 59252.2124933(16) & 0.1677075 \\ [0.8ex]
ZTF\,J070458.08$-$020103.3 & $74.3^{+0.4}_{-0.2}$ & $1.079^{+0.006}_{-0.010}$ & $0.052^{+0.011}_{-0.013}$ & $3.715^{+0.076}_{-0.075}$ & $3.643\pm0.088$ & 59253.2216462(43) & 0.1413708 \\ [0.8ex]
ZTF\,J071759.04$+$113630.2 & $84.9^{+0.4}_{-0.3}$ & $2.326^{+0.027}_{-0.030}$ & $0.018^{+0.017}_{-0.012}$ & $2.812^{+0.065}_{-0.072}$ & $2.74\pm0.13$ & 59251.1312794(93) & 0.4527638 \\ [0.8ex]
ZTF\,J071843.68$-$085232.1 & $84.6^{+0.7}_{-0.7}$ & $1.563^{+0.014}_{-0.013}$ & $0.064^{+0.017}_{-0.019}$ & $2.157^{+0.062}_{-0.058}$ & $2.39\pm0.22$ & 59283.1026109(12) & 0.2158113 \\ [0.8ex]
ZTF\,J080441.95$-$021545.7 & $85.3^{+0.1}_{-0.1}$ & $0.889^{+0.007}_{-0.004}$ & $0.027^{+0.015}_{-0.014}$ & $5.631^{+0.092}_{-0.089}$ & $5.47\pm0.11$ & 59646.09050723(97) & 0.1209762 \\ [0.8ex]
ZTF\,J080542.98$-$143036.3 & $81.0^{+1.0}_{-0.7}$ & $1.260^{+0.015}_{-0.019}$ & $0.010^{+0.012}_{-0.008}$ & $1.102^{+0.034}_{-0.039}$ & $1.39\pm0.16$ & 59646.18599526(74) & 0.1981669 \\ [0.8ex]
ZTF\,J094826.35$+$253810.6 & $79.9^{+0.6}_{-0.6}$ & $1.003^{+0.017}_{-0.017}$ & $0.018^{+0.008}_{-0.009}$ & $2.911^{+0.116}_{-0.108}$ & $2.94\pm0.26$ & 59239.2668295(95) & 0.1418270 \\ [0.8ex]
ZTF\,J102254.00$-$080327.3 & $76.9^{+0.6}_{-0.6}$ & $1.592^{+0.024}_{-0.023}$ & $0.019^{+0.013}_{-0.012}$ & $5.750^{+0.158}_{-0.160}$ & $5.58\pm0.21$ & 59280.2576703(43) & 0.2314179 \\ [0.8ex]
ZTF\,J102653.47$-$101330.3 & $87.4^{+1.6}_{-1.5}$ & $0.677^{+0.008}_{-0.006}$ & $0.033^{+0.010}_{-0.010}$ & $2.027^{+0.075}_{-0.063}$ & $1.65\pm0.19$ & 59237.2453759(14) & 0.0929868 \\ [0.8ex]
ZTF\,J103448.82$+$005201.9 & $87.8^{+0.3}_{-0.2}$ & $0.688^{+0.003}_{-0.003}$ & $0.031^{+0.016}_{-0.015}$ & $3.551^{+0.138}_{-0.138}$ & $3.20\pm0.28$ & 59253.2517553(11) & 0.0915591 \\ [0.8ex]
ZTF\,J104906.96$-$175530.7 & $88.6^{+1.1}_{-1.1}$ & $1.407^{+0.012}_{-0.006}$ & $0.029^{+0.006}_{-0.008}$ & $2.579^{+0.070}_{-0.070}$ & $2.47\pm0.17$ & 59238.3654607(11) & 0.2447332 \\ [0.8ex]
ZTF\,J122009.98$+$082155.0 & $87.5^{+0.2}_{-0.2}$ & $4.592^{+0.052}_{-0.056}$ & $0.024^{+0.004}_{-0.008}$ & $3.874^{+0.093}_{-0.106}$ & $3.53\pm0.17$ & 59252.2559984(25) & 1.2329254 \\ [0.8ex]
ZTF\,J125620.57$+$211725.8 & $89.8^{+0.2}_{-0.2}$ & $2.374^{+0.013}_{-0.012}$ & $0.005^{+0.006}_{-0.004}$ & $22.221^{+0.095}_{-0.094}$ & $22.171\pm0.096$ & 59641.3540758(33) & 0.5560572 \\ [0.8ex]
ZTF\,J130228.34$-$003200.2 & $86.0^{+0.5}_{-0.6}$ & $1.268^{+0.012}_{-0.008}$ & $0.016^{+0.013}_{-0.011}$ & $8.554^{+0.071}_{-0.068}$ & $8.555\pm0.073$ & 59252.387889(43) & 0.1661310 \\ [0.8ex]
ZTF\,J134151.70$-$062613.9 & $86.8^{+2.4}_{-2.7}$ & $0.764^{+0.018}_{-0.017}$ & $0.030^{+0.009}_{-0.009}$ & $0.894^{+0.089}_{-0.093}$ & $0.97\pm0.12$ & 59237.3073649(28) & 0.0969505 \\ [0.8ex]
ZTF\,J140036.65$+$081447.4 & $89.2^{+0.6}_{-0.7}$ & $1.589^{+0.006}_{-0.005}$ & $0.005^{+0.008}_{-0.004}$ & $2.155^{+0.070}_{-0.075}$ & $1.58\pm0.30$ & 59253.2966645(14) & 0.2602766 \\ [0.8ex]
ZTF\,J140423.86$+$065557.7 & $84.5^{+1.0}_{-0.9}$ & $1.342^{+0.010}_{-0.009}$ & $0.025^{+0.007}_{-0.009}$ & $2.538^{+0.059}_{-0.057}$ & $2.24\pm0.14$ & 59239.3665054(12) & 0.1683096 \\ [0.8ex]
ZTF\,J140537.34$+$103919.0 & $88.5^{+0.4}_{-0.3}$ & $1.389^{+0.009}_{-0.009}$ & $0.016^{+0.010}_{-0.009}$ & $0.752^{+0.031}_{-0.024}$ & $0.78\pm0.26$ & 59251.334651(12) & 0.2714122 \\ [0.8ex]
ZTF\,J140702.57$+$211559.7 & $86.5^{+2.4}_{-2.0}$ & $1.008^{+0.016}_{-0.011}$ & $0.051^{+0.010}_{-0.013}$ & $4.077^{+0.072}_{-0.070}$ & $4.079\pm0.091$ & 59643.3349542(63) & 0.1432802 \\ [0.8ex]
ZTF\,J145819.54$+$131326.7 & $86.8^{+0.1}_{-0.2}$ & $0.742^{+0.004}_{-0.003}$ & $0.025^{+0.009}_{-0.012}$ & $5.067^{+0.152}_{-0.154}$ & $4.86\pm0.21$ & 59252.3531663(17) & 0.0920516 \\ [0.8ex]
ZTF\,J162644.18$-$101854.3 & $88.7^{+0.9}_{-1.1}$ & $1.503^{+0.013}_{-0.010}$ & $0.291^{+0.007}_{-0.013}$ & $1.733^{+0.087}_{-0.085}$ & $1.91\pm0.20$ & 59253.3679108(15) & 0.2530067 \\ [0.8ex]
ZTF\,J163421.00$-$271321.7 & $80.6^{+2.1}_{-1.5}$ & $0.637^{+0.020}_{-0.028}$ & $0.182^{+0.029}_{-0.026}$ & $4.127^{+0.238}_{-0.230}$ & $4.24\pm0.26$ & 59253.3310632(36) & 0.0780396 \\ [0.8ex]
ZTF\,J164441.18$+$243428.2 & $80.3^{+0.6}_{-0.7}$ & $0.614^{+0.011}_{-0.011}$ & $0.031^{+0.017}_{-0.015}$ & $2.197^{+0.087}_{-0.086}$ & $2.43\pm0.22$ & 59283.3945858(11) & 0.0801054 \\ [0.8ex]
ZTF\,J180256.45$-$005458.3 & $84.2^{+0.3}_{-0.3}$ & $1.485^{+0.020}_{-0.023}$ & $0.114^{+0.028}_{-0.024}$ & $4.700^{+0.077}_{-0.101}$ & $4.38\pm0.15$ & 59646.3585478(21) & 0.2690033 \\ [0.8ex]
ZTF\,J182848.77$+$230838.0 & $88.7^{+0.1}_{-0.3}$ & $0.852^{+0.005}_{-0.002}$ & $0.088^{+0.014}_{-0.018}$ & $4.955^{+0.079}_{-0.077}$ & $4.914\pm0.097$ & 59695.37741036(84) & 0.1120067 \\ [0.8ex]
ZTF\,J195456.71$+$101937.5 & $84.5^{+0.7}_{-0.8}$ & $1.901^{+0.020}_{-0.016}$ & $0.078^{+0.022}_{-0.023}$ & $3.495^{+0.051}_{-0.050}$ & $3.449\pm0.057$ & 59697.3389707(12) & 0.3102884 \\ [0.8ex]
\hline
\end{tabular}
\end{table*}

\section{Discussion}

\subsection{Comparison with previous parameters}

\begin{figure*}
 \includegraphics[width=\textwidth]{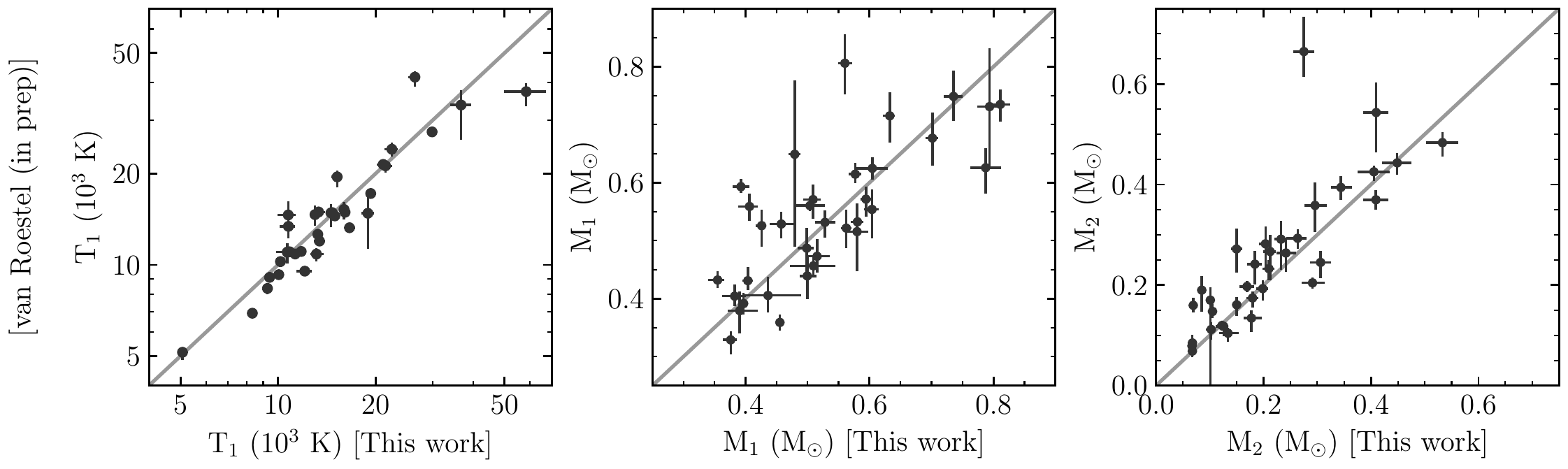}
 \caption{Comparison of our parameters from the NTT-ULTRACAM photometry against the initial parameters of van Roestel et al. (in prep) from ZTF photometry.}
 \label{fig:comparison}
\end{figure*}

Initial parameter estimates for these systems were made by fitting the ZTF time-series photometry alongside photometric measurements from other surveys, where available, covering a wide wavelength range (van Roestel et al. in prep).
Comparing our parameters determined from the three-band eclipse photometry against these initial estimates demonstrates general agreement between the two methods (\autoref{fig:comparison}).
The WD temperatures, in particular, show excellent agreement but there are some significant differences in the measured masses for certain systems.
This may be due, in part, to the survey SED data used by van Roestel et al. (in prep) being taken at a range of different orbital phases and therefore suffering from increased systematics due to ellipsoidal modulation or reflection effect.
As the method we have used in this work has previously been shown to retrieve accurate parameters \citep{Brown2022}, this may imply a slight underestimation in the uncertainties determined by combining SED fitting with ZTF photometry.
The parameters determined using the high-speed eclipse photometry are typically more precise than those measured by van Roestel et al. (in prep).
This is most apparent for the primary and secondary masses with a median uncertainty in the WD mass from the ULTRACAM photometry of 2.6~per~cent, and 7.2~per~cent for the secondary mass.
These values are 6.0~per~cent and 13.7~per~cent respectively from the ZTF photometry for the same systems and so the ULTRACAM measurements are typically a factor of 2 more precise.
This is likely due to the high time resolution of the ULTRACAM photometry, enabling the duration of the eclipse as well as the ingress and egress to be measured very precisely.

In addition to the initial parameter estimates discussed above, two of the systems fit in this work have been included in previously published analyses -- ZTF~J125620.57$+$211725.8 and ZTF~J164441.18$+$243428.2.
Comparisons with these previous works are made below.

\subsubsection{ZTF~J125620.57$+$211725.8}
ZTF~J125620.57$+$211725.8 was previously fitted by \citet{Rebassa-Mansergas2021}, using the Virtual Observatory SED Analyser (VOSA) to fit the available survey photometry.
Out of the 112 systems that they analysed, 13 systems were determined to possess a WD with a mass below $0.2~\rm{M_{\odot}}$.
It is not known how such low mass WDs could form in PCEBs with low-mass main sequence companions -- with any mass transfer initiating a common envelope phase in which the envelope would most likely not gain sufficient energy to be ejected, leading to a merger scenario \citep{Rebassa-Mansergas2021}.
This system is -- as far as we know -- the only one of these 13 systems that eclipses, enabling a valuable check on the system parameters.
Our fit to the eclipse photometry determines the WD mass to be $0.48\pm0.01~\mathrm{M_{\odot}}$, discrepant with the $0.155\pm0.02~\mathrm{M_{\odot}}$ obtained from VOSA by over $14\sigma$.
We encourage spectroscopic follow-up of this system in order to determine the cause of this large discrepancy and the true WD mass.

\subsubsection{ZTF~J164441.18$+$243428.2}
ZTF~J164441.18$+$243428.2 was one of the four deeply eclipsing PCEBs found and fitted by \citet{Kosakowski2022}.
For this target in particular they did not detect the eclipse minimum and so their parameters from the light curve fit represent limits rather than specific values.
As would be expected, our light curve fit to the ULTRACAM photometry is consistent with these parameter limits.
As well as fitting the eclipse light curve \citet{Kosakowski2022} performed a spectroscopic fit to the WD, determining the effective temperature, surface gravity, and mass (determined from the surface gravity using CO-core composition models).
From our fit to the ULTRACAM photometry we find an effective temperature of $13270\pm490~\rm{K}$, cooler than the $14900\pm760~\rm{K}$ determined by their spectroscopic fit but still consistent to within $2\sigma$.
For the WD mass there is a little more deviation, with our fit finding a WD mass of $0.38\pm0.02~\mathrm{M_{\odot}}$, $2.3\sigma$ below the $0.55\pm0.07~\mathrm{M_{\odot}}$ found from their spectroscopic fit and suggesting a He-core composition rather than a CO-core.
For the companion, \citet{Kosakowski2022} estimate a mass of $0.084\pm0.004~\mathrm{M_{\odot}}$ by fitting the Pan-STARRS SED with a composite model, placing it close to the hydrogen-burning limit.
We find a higher mass of $0.103\pm0.009~\mathrm{M_{\odot}}$ from our light curve fit taking it into more typically stellar territory.
Again though, these two values are consistent to within $2\sigma$.
Overall, our fit to the ULTRACAM photometry is fully consistent with their light curve fit and consistent with their spectroscopic and Pan-STARRS SED fits at around the $2\sigma$ level.

\subsection{Brown dwarf companions}

WDs with brown dwarf companions are rare, with around 0.5 per cent of WDs expected to have substellar partners \citep{Steele2011}.
Eclipsing examples are, predictably, even rarer with only four systems currently confirmed \citep{Beuermann2013, Littlefair2014, Parsons2017b, Casewell2020a, vanRoestel2021}.
These eclipsing WD-brown dwarf binaries are valuable as they are one of the few places where both the brown dwarf's radii and mass can be measured precisely and are therefore important benchmarks for brown dwarf models.
Additionally, as some of the lowest mass objects thought to survive the common-envelope \citep{Casewell2018b}, brown dwarfs in PCEBs occupy an important area of the parameter space when studying common-envelope evolution, with the study of the common-envelope phase in this low-mass regime having implications for systems with planetary mass companions \citep{Vanderburg2020}.

\begin{figure}
 \includegraphics[width=\columnwidth]{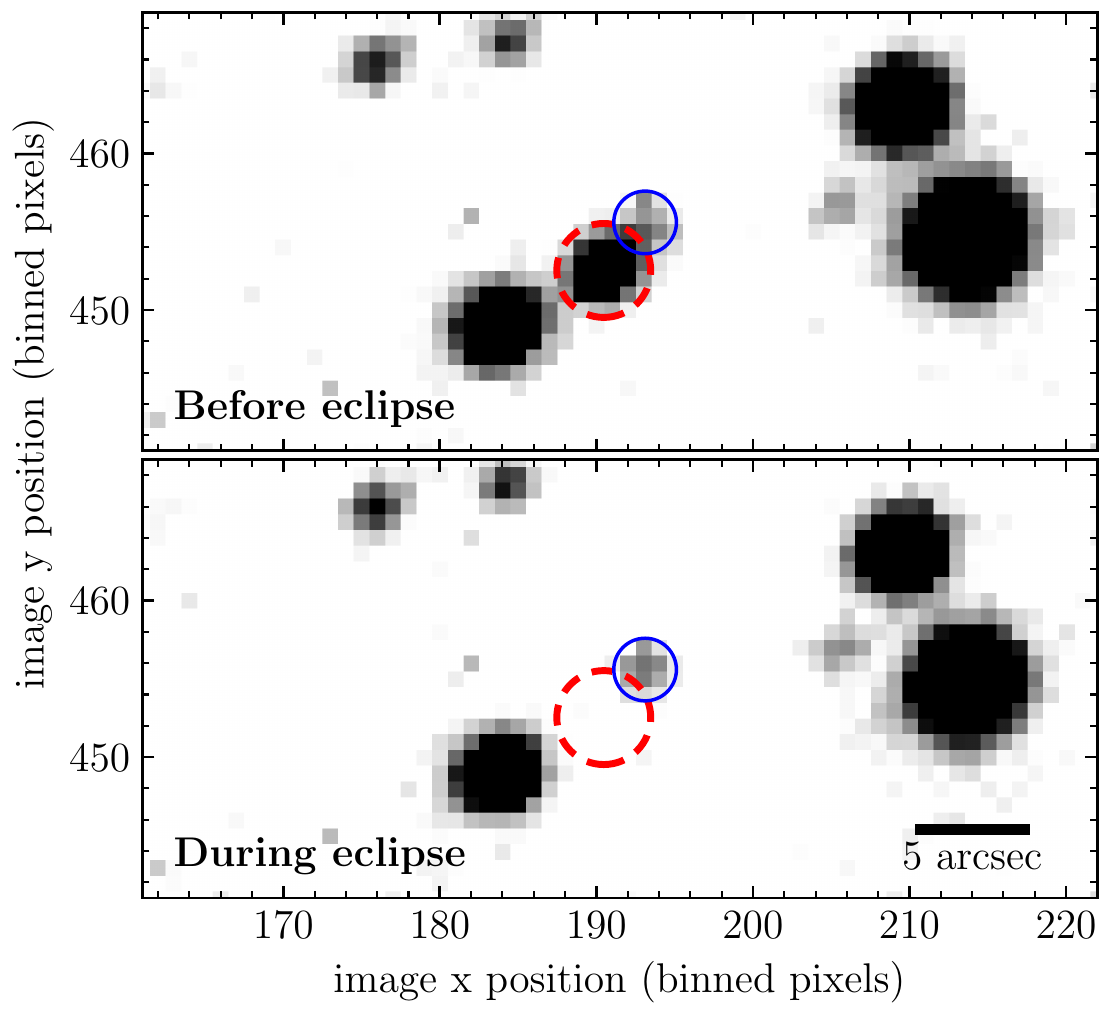}
 \caption{Stacked images of ZTF\,J1828$+$2308 taken with ULTRACAM in the $i_{s}$ filter before and during the eclipse. The red dashed aperture shows the location of ZTF\,J1828$+$2308 itself while the solid blue aperture shows the fainter background source 2.79" away (Gaia\,DR3\,4529477702982880512) that is marginally affecting our in-eclipse photometry.}
 \label{fig:ZTFJ1828_image}
\end{figure}

In our ULTRACAM follow-up we have found four systems so far that our light curve fits suggest as having brown dwarf companions.
These are ZTF\,J080441.95$-$021545.7, ZTF\,J103448.82$+$005201.9, ZTF\,J145819.54$+$131326.7, and ZTF\,J182848.77$+$230838.0.
As our mass-radius relation for M~dwarfs \citep{Brown2022} is horizontal below $0.07~\rm{M_{\odot}}$ -- and therefore uninformative in this regime -- the best fit secondary masses can only be regarded as upper limits.
Additionally, as none of the secondaries for these systems are detected in-eclipse, only an upper limit can be given for their effective temperatures. One of these systems, ZTF\,J182848.77$+$230838.0, has a high secondary temperature for a brown dwarf.
In order to rule out problems with the photometry, we stack the in-eclipse images (\autoref{fig:ZTFJ1828_image}).
This reveals a faint ($G=20.88~\rm{mag}$) source 2.79~arcsec away from the target which results in an erroneous slight `detection' in eclipse and therefore a higher than expected temperature.
The true upper limit for the secondary temperature will be lower than given by our fit.

\begin{figure}
 \includegraphics[width=\columnwidth]{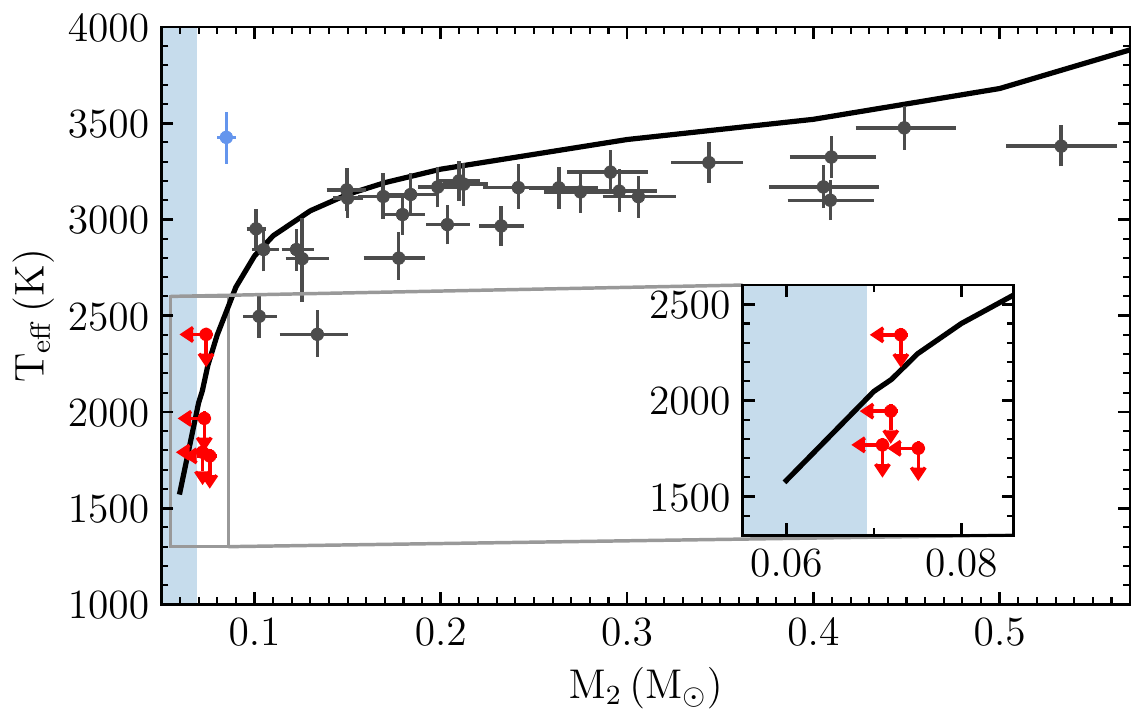}
 \caption{Measured masses and effective temperatures of the M~dwarf components with an inset plot zoomed in around the brown dwarfs (which are shown in red). The solid black line shows the 1\,Gyr track from \citet{Baraffe2015} and the shaded blue area denotes the region where our mass-radius relation is horizontal (i.e. the radius is constant in this mass range). For the brown dwarfs we plot the masses and temperatures as upper limits centred on the $84^{\rm{th}}$ percentile of the fit. The blue point denotes ZTF~J1405+1039 which has a best-fit secondary temperature that is much hotter than expected for its mass.}
 \label{fig:t2_m2}
\end{figure}

\begin{figure}
 \includegraphics[width=\columnwidth]{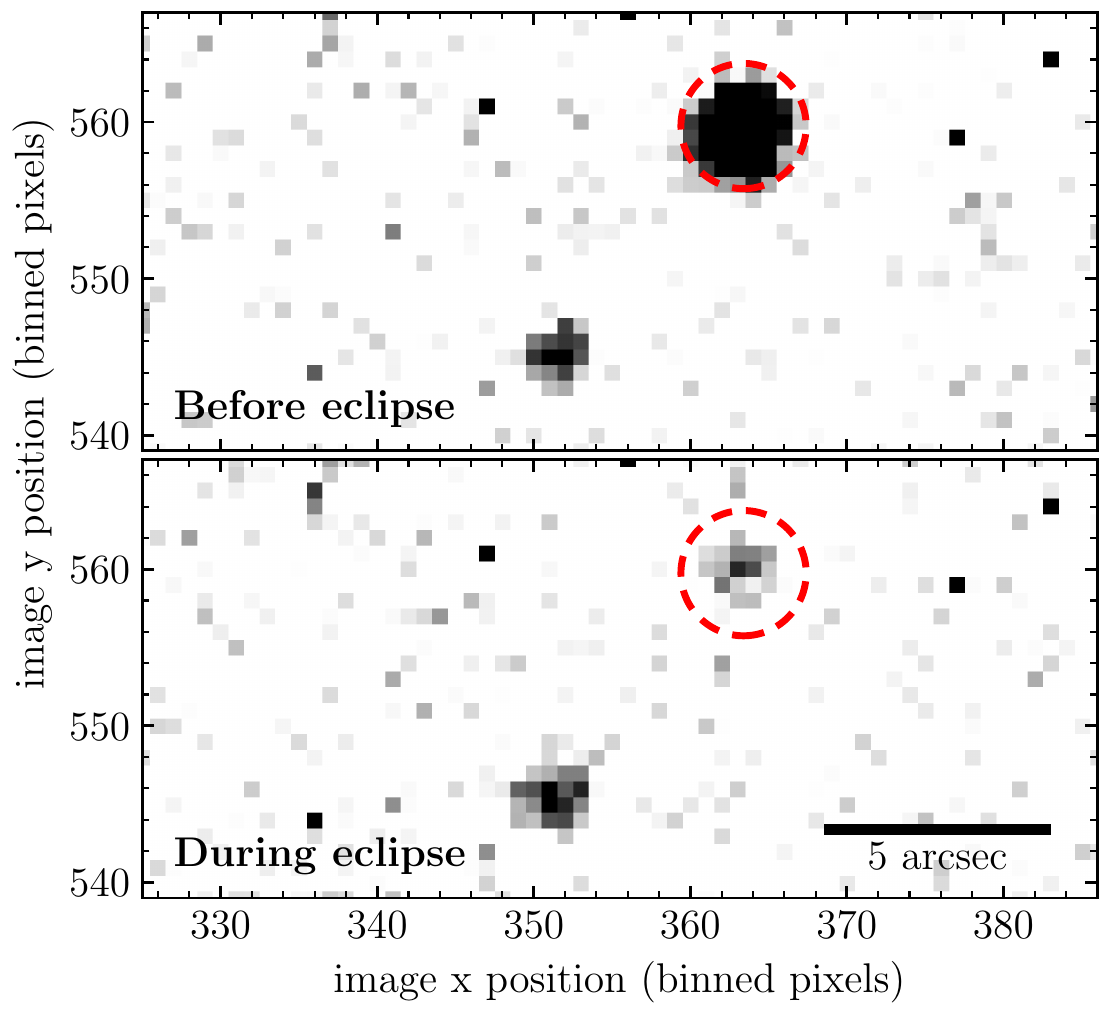}
 \caption{Stacked images of ZTF\,J1405$+$1039 taken with ULTRACAM in the $i_{s}$ filter before and during the eclipse. The red dashed aperture shows the location of ZTF\,J1405$+$1039. It is clear that the source is still detected in-eclipse.}
 \label{fig:ZTFJ1405_image}
\end{figure}

In addition to these four systems with sub-stellar companions, we have measured one system with a companion mass just above the hydrogen-burning limit, ZTF\,J140537.34$+$103919.0, hereafter ZTF\,J1405$+$1039.
The best-fit parameters for this system suggest that the secondary is significantly hotter than would be expected for its mass (shown as the blue point in \autoref{fig:t2_m2}).
Again, we stack the in-eclipse images to rule out problems in the photometry (\autoref{fig:ZTFJ1405_image}), demonstrating that the source is indeed detected in-eclipse.
We believe that the most likely explanation for this is that ZTF\,J1405$+$1039 is actually a triple system, with a tertiary companion contributing a significant fraction of the in-eclipse flux.

\subsection{ZZ Ceti WDs}
ZZ Cetis are pulsating WDs, possessing hydrogen atmospheres and pulsation periods ranging from tens of seconds to tens of minutes \citep{Fontaine2008, Winget2008, Romero2022a}.
The presence of pulsations enable asteroseismological analyses to be performed, providing insight into the internal structure of the WD which is otherwise concealed by their highly stratified nature.
In PCEBs, the possibility of measuring the internal structure of the WD is especially interesting as it can reveal how the WD itself is affected by the common-envelope phase \citep{Hermes2015}.
Previously, only one ZZ Ceti WD in a detached eclipsing binary was known \citep{Parsons2020}. This system is a double WD binary, however, and as such its evolutionary history is less well defined, with the number of common-envelope events it has passed through being uncertain.
ZZ Cetis found in WD-main sequence PCEBs do not have this problem with their evolutionary past known to comprise of a single common-envelope phase. These systems are therefore potentially very interesting systems to find.
Currently there is one known ZZ Ceti WD in a detached, albeit not eclipsing, PCEB \citep{Pyrzas2015}.
Although this is an important find, \citet{Hermes2015} noted that there were a lot of free parameters, limiting the precision of the asteroseismological analysis.
Eclipsing examples of such systems would reduce these free parameters and enable a more precise analysis.

\begin{figure}
 \includegraphics[width=\columnwidth]{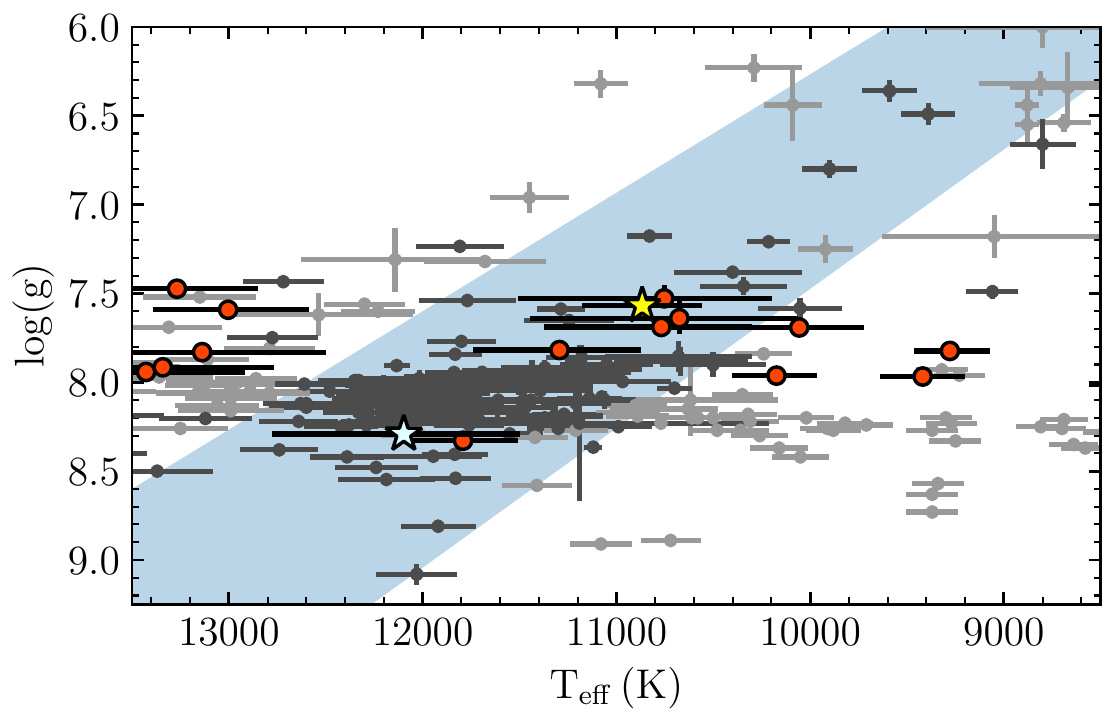}
 \caption{The ZZ Ceti instability strip (blue region) with known pulsating (dark grey) and non-pulsating (light grey) WDs from \citet{Gianninas2011, Steinfadt2012, Hermes2012, Hermes2013a, Hermes2013b, Hermes2013c, Romero2022a}. Points in red show the measured parameters of the WD components of binaries fit in this work, with the confirmed pulsators, ZTF~J1407+2115 and ZTF~J0528+2156, shown by the yellow and cyan stars respectively.}
 \label{fig:instability_strip}
\end{figure}

Comparing our best fit parameters for the WD components to the ZZ Ceti instability strip (\autoref{fig:instability_strip}), we find that eight of our systems have WDs that lie within $1\sigma$ of the instability strip (shown in \autoref{tab:pulsating}).
Closer inspection of their light curves do not reveal any clear photometric variability indicative of pulsations in six of the systems, however, the out-of-eclipse data for many of these systems is typically less than 30 minutes and so is not enough to rule out pulsations either. Although the WD temperatures are not necessarily precise enough to say with certainty whether a particular WD lies within the instability strip or not.
Of these eight systems with WDs that lie in the instability strip, we have found two that show clear variability due to pulsations. These represent the first two ZZ Ceti WDs found in eclipsing WD+dM PCEBs.

\begin{table}
    \centering
    \begin{tabular}{ccccc}
        \hline
        Target & RA & Dec & G & Pulsating \\
        \hline
        ZTF\,J0519$+$0925 & 05:19:02.1 & +09:25:26.38 & 19.0 & Candidate \\
        ZTF\,J0528$+$2156 & 05:28:48.2 & +21:56:28.94 & 17.7 & Confirmed \\
        ZTF\,J0948$+$2538 & 09:48:26.4 & +25:38:10.68 & 18.7 & Candidate \\
        ZTF\,J1034$+$0052 & 10:34:48.8 & +00:52:01.69 & 19.0 & Candidate \\
        ZTF\,J1302$-$0032 & 13:02:28.3 & -00:32:00.11 & 16.8 & Candidate \\
        ZTF\,J1407$+$2115 & 14:07:02.6 & +21:15:59.75 & 17.4 & Confirmed \\
        ZTF\,J1634$-$2713 & 16:34:21.0 & -27:13:21.54 & 18.8 & Candidate \\
        ZTF\,J1802$-$0054 & 18:02:56.4 & -00:54:58.47 & 18.0 & Candidate \\
        \hline
    \end{tabular}
    \caption{eclipsing PCEBs with -- either confirmed or candidate -- ZZ Ceti WDs}
    \label{tab:pulsating}
\end{table}

\subsubsection{ZTF~J1407$+$2115}
\begin{figure}
 \includegraphics[width=\columnwidth]{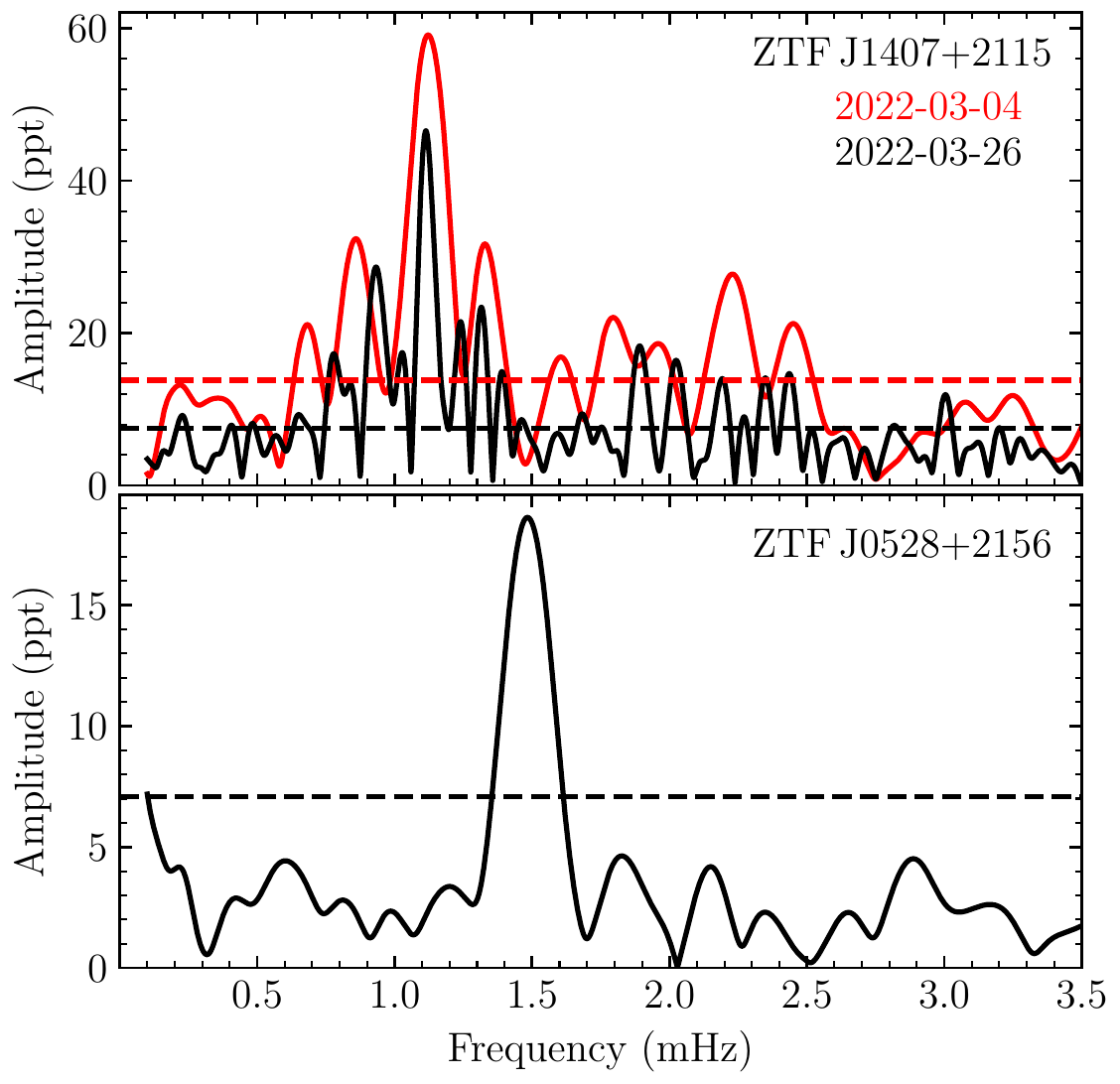}
 \caption{Lomb-Scargle periodograms (shown in parts per thousand relative to the flux of the WD) of the ULTRACAM $g_{s}$ light curves of ZTF~J1407$+$2115 and ZTF~J0528$+$2156 with their respective eclipse light curve models subtracted. Horizontal dashed lines show the $3\sigma$ significance levels calculated using the bootstrapping method described by \citet[Section 4.1]{Greiss2014}.}
 \label{fig:periodograms}
\end{figure}

ZTF~J140702.56$+$211559.7, hereafter ZTF~J1407$+$2115, was first observed with ULTRACAM in February 2021.
Unusual out-of eclipse variation was noticed but the data taken in this run was insufficient to confirm pulsations.
We observed ZTF~J1407+2115 again for 1\,h on the $2^{\rm{nd}}$ of March 2022, detecting 3 clear pulsations and confirming it as the first eclipsing detached PCEB containing a ZZ Ceti WD.
With this confirmation, we observed ZTF~J1407+2115 in two long observing runs on the $4^{\rm{th}}$ and $26^{\rm{th}}$ of March 2022 using the $u_{s}\,g_{s}\,i_{s}$ and $u_{s}\,g_{s}\,r_{s}$ filters and lasting $\sim2\,\rm{h}$ and $\sim5\,\rm{h}$ respectively (Lomb-scargle periodograms of these two long runs are shown in \autoref{fig:periodograms}).
It is the photometry from the long observing run on the $4^{\rm{th}}$ of March that we use to fit the system parameters.
We choose this observation primarily for consistency with the modelling performed on the other systems in this work, but also as the wider wavelength range provided by the $i_{s}$-band strengthens the constraints on the WD temperature. Additionally, chromospheric variability in the H$\alpha$ feature can lead to higher scatter of M~dwarf fluxes in the $r_{s}$-band.

\begin{figure*}
 \includegraphics[width=\textwidth]{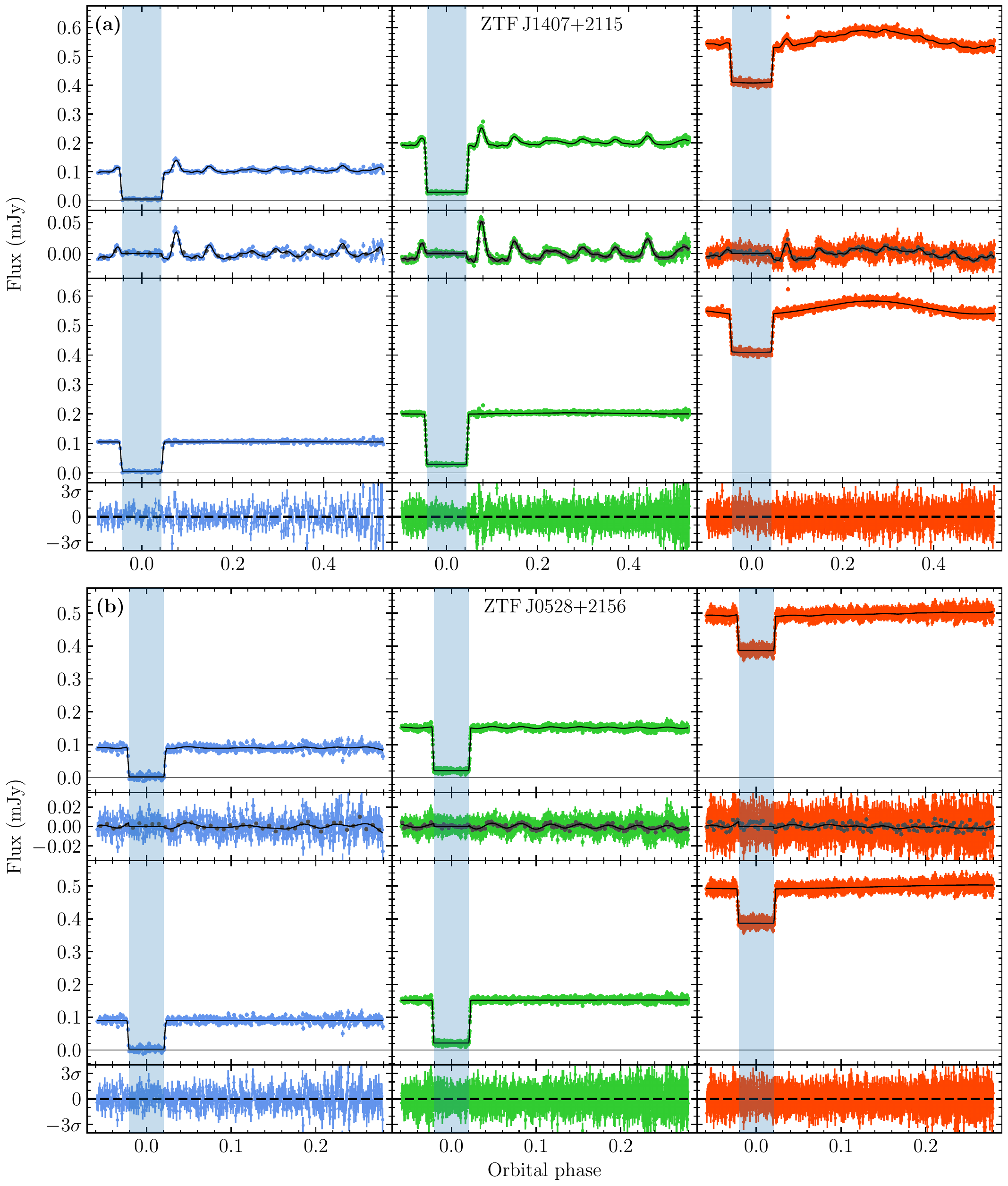}
 \caption{ULTRACAM $u_{s}\,g_{s}\,i_{s}$ light curves of ZTF~J1407$+$2115 (\textbf{a}) and ZTF~J0528$+$2156 (\textbf{b}). The top row of each plot shows the observed light curve (coloured points) with the combined eclipse plus mean Gaussian process pulsation model (black line). The second row shows the observed light curve with the eclipse model subtracted (coloured points) as well as the same data binned up by a factor of ten (dark grey points) with the mean Gaussian process model (black line). The third row shows the observed light curve with the mean Gaussian process subtracted with the black line showing the eclipse model. The bottom row shows the residuals of the full light curve model. The filled region shows the phase range where the Gaussian process is switched off (between the second and third eclipse contact points).}
 \label{fig:GP_LC}
\end{figure*}

In order to fit the eclipse photometry of this system, the pulsations need to be included in the light curve model to prevent them introducing large systematic errors in the best-fit parameters.
We do this using a Gaussian process (GP) implemented through the \textsc{python} package, \textsc{george}\footnote{\url{https://george.readthedocs.io/en/latest/}} \citep{Ambikasaran2015}.
The GP is applied to the residuals of the \textsc{pylcurve} model at each MCMC walker position, with the posterior log probability calculated as the sum of the GP marginalised log likelihood, the log likelihood from comparing the model WD SED with the measured eclipse depths, and the log priors (parallax and interstellar reddening).
We use the \texttt{ExpSquaredKernel}, defined by an amplitude, temperature, and scale-length, with the temperature scaling the pulsation amplitude between the light curves in different filters according to a blackbody law.
These three GP parameters are included as free parameters in our fit.
We switch the GP off between the second and third contact points where the WD is totally eclipsed by its M~dwarf companion, with the contact points being calculated for every walker position.
We then use \textsc{emcee} \citep{Foreman-Mackey2013} to sample from the posterior probability distribution and determine the best-fit parameters.
This best-fit model is shown in \autoref{fig:GP_LC}.
We find the WD to have an effective temperature of $10\,900\pm300~\rm{K}$ and a mass of $0.41\pm0.01~\rm{M_{\odot}}$, suggesting a core composed primarily of helium.
This mass and temperature corresponds to a surface gravity of $7.57\pm0.04~\rm{dex}$, placing it in a relatively sparsely sampled region in the middle of the instability strip (\autoref{fig:instability_strip}).

We subtract our best-fit eclipse light curve model from the $g_{s}$-band photometry of the longer run on the $26^{\rm{th}}$ of March, leaving just the pulsation signal. Running a periodogram on this determines the main pulsation mode to have a frequency of 1.11~mHz (898~s) with an amplitude of around 47~parts per thousand (ppt) (\autoref{fig:periodograms}). We calculate the $3\sigma$ significance threshold to be 8~ppt following the method of \citet{Greiss2014}, shuffling the flux values 10\,000 times and taking the amplitude of the $99.7^{\rm{th}}$ percentile highest peak.

\subsubsection{ZTF~J0528$+$2156}
ZTF~J052848.24$+$215629.0, hereafter ZTF~J0528$+$2156, was first observed with ULTRACAM in February 2021.
Attempts at fitting the eclipse light curve showed some possible structure in the residuals, prompting us to observe it again to search for pulsations.
We observed ZTF~J0528$+$2156 again on the $18^{\rm{th}}$ of December 2022 for 1.8~h, detecting pulsations with a period of around 11~minutes and amplitude of around 5~per~cent.

We fit the ULTRACAM photometry in the same way as for ZTF~J1407$+$2115 -- using a Gaussian process to model the pulsations.
We find the WD to have an effective temperature of $11\,900\pm600~\rm{K}$ and a mass of $0.78\pm0.02~\rm{M_{\odot}}$, corresponding to a surface gravity of $8.27\pm0.044~\rm{dex}$ and placing it comfortably within the instability strip (\autoref{fig:instability_strip}).
Computing the periodogram of the residuals of the eclipse light curve model in the same way as for ZTF~J1407$+$2115, we find the main mode to have a frequency of 1.5~mHz (670~s) and amplitude of around 19~ppt with a $3\sigma$ significance threshold of 7~ppt (\autoref{fig:periodograms}).

\subsection{Magnetic WDs}
\label{sec:magnetic_wd}
Around 36 per cent of WDs in cataclysmic variables (CVs) are observed to be strongly magnetic \citep{Pala2020}.
This is in stark contrast with their progenitor population -- the detached PCEBs -- of which only a handful possess WDs with strong magnetic fields.
\citet{Schreiber2021} propose an evolutionary channel between the magnetic CVs and the detached magnetic population to explain this discrepancy.
This relies on a rotation-driven dynamo in which a crystallising WD, spun up due to accretion during the CV phase, can generate the strong magnetic fields that we observe in CVs.
Interactions between the newly-formed magnetic field of the WD and the magnetic field of the M~dwarf then act to detach the binary, halting mass transfer and causing the binary to appear as a strongly magnetic detached PCEB for a period of time before angular momentum loss due to magnetic braking and gravitational wave radiation brings the two stars back into a mass-transferring state as a polar or intermediate polar.

A test of this model was performed by \citet{Parsons2021}, using spectroscopic observations of detached magnetic PCEBs to constrain their evolutionary history, attempting to assess whether or not they are consistent with having undergone a mass-transferring phase in the past. All systems studied were found to be consistent with a previous CV phase but spectroscopic observations alone were not powerful enough to draw strong conclusions. More powerful constraints can be made if such systems are found to be eclipsing, enabling more precise measurements to be made from the eclipse photometry and therefore a robust test of the model.

\begin{figure*}
 \includegraphics[width=\textwidth]{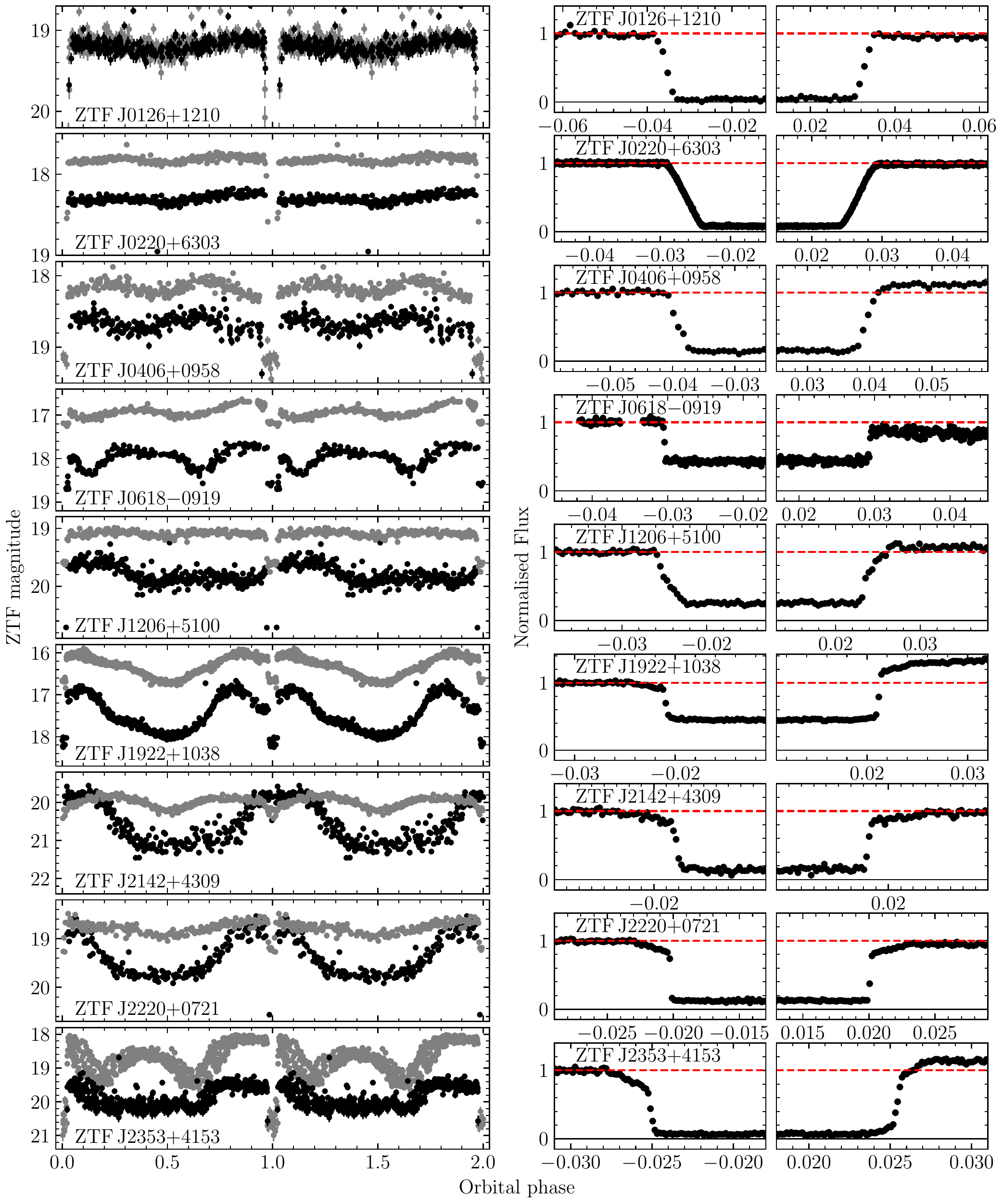}
 \caption{{\it left}: ZTF g-band (black) and r-band (grey) light curves of the 6 new confirmed, and 3 new candidate eclipsing PCEBs with magnetic WDs. All show out-of-eclipse variation inconsistent with reflection effect or ellipsoidal modulation in at least one filter. Some light curves have been binned for clarity. {\it right}:~Normalised ULTRACAM/HiPERCAM $g_{s}$-band primary eclipse light curves (zoomed in on the ingress and egress) of the 6 new confirmed, and 3 new candidate eclipsing PCEBs with magnetic WDs. The solid grey line shows a flux of zero while the red dashed line shows the mean flux of the first 10 points shown.}
 \label{fig:ztf_ucam_mag}
\end{figure*}

As part of our follow-up program we have discovered 6 new eclipsing PCEBs (\autoref{tab:magnetic}) that we have confirmed from our high-speed photometry as having magnetic WDs -- showing clear evidence of a bright magnetic pole in the eclipse ingress/egress, with one previously known as a magnetic system but not known to be eclipsing. We have additionally found 3 candidate systems that show out-of-eclipse variation that disappears when the WD is eclipsed but for which the ingress/egress of the eclipse do not confirm a bright magnetic pole. These systems have been found by searching for unusual out-of-eclipse variation in their ZTF light curves (\autoref{fig:ztf_ucam_mag}), inconsistent with the ellipsoidal modulation or reflection effect that is common in PCEBs. This unusual out-of-eclipse variability was noted in the pre-intermediate polar, SDSS\,J0303$+$0054, \citep{Parsons2013b} and is due to additional emission in the form of cyclotron radiation from the magnetic poles of the WD. The effect of the cyclotron emission on the eclipse profiles -- introducing steps in the ingress and egress due to the eclipse of the small, bright magnetic pole  (\autoref{fig:ztf_ucam_mag}) -- makes the light curves of the magnetic systems more complicated to fit and so the analysis of these systems will be the subject of a future paper.

\begin{table}
    \centering
    \begin{tabular}{ccccc}
        \hline
        Target & RA & Dec & G & Magnetic \\
        \hline
        ZTF\,J0126$+$1210 & 01:26:07.8 & $+$12:10:49.14 & 18.8 & Candidate \\
        ZTF\,J0220$+$6303 & 02:20:04.6 & $+$63:03:59.63 & 17.4 & Candidate \\
        ZTF\,J0406$+$0958 & 04:06:27.2 & $+$09:58:26.97 & 18.1 & Candidate \\
        ZTF\,J0618$-$0919 & 06:18:09.9 & $-$09:19:04.28 & 16.5 & Confirmed \\
        ZTF\,J1206$+$5100 & 12:06:15.7 & $+$51:00:46.77 & 18.6 & Confirmed \\
        ZTF\,J1922$+$1038 & 19:22:15.3 & $+$10:38:38.13 & 16.0 & Confirmed \\
        ZTF\,J2142$+$4309 & 21:42:32.0 & $+$43:09:28.97 & 19.3 & Confirmed \\
        ZTF\,J2220$+$0721 & 22:20:07.5 & $+$07:21:29.74 & 18.3 & Confirmed \\
        ZTF\,J2353$+$4153 & 23:53:55.0 & $+$41:53:04.40 & 18.7 & Confirmed \\
        \hline
    \end{tabular}
    \caption{eclipsing PCEBs with -- either confirmed or candidate -- magnetic WD components. }
    \label{tab:magnetic}
\end{table}

\section{Conclusions}

Through our dedicated program of high-speed photometric follow-up we have obtained multi-band eclipse light curves for 43 new PCEBs found using ZTF.
We have characterized 34 of these systems from the eclipse light curves alone -- finding four that contain sub-stellar companions, doubling the number of eclipsing examples known, and two with pulsating WDs representing the first ZZ Ceti WDs known in eclipsing WD+dM binaries.
Of the remaining nine systems, we have found six to contain strongly magnetic WDs from their eclipse photometry with three further candidates.
These will be invaluable to the study of magnetic field generation in binary WDs.
Our results demonstrate that a photometric approach to the follow-up of eclipsing systems can effectively discern interesting sub-types of PCEBs, including those that would be otherwise missed by spectroscopic follow-up.

\section*{Acknowledgements}

SGP acknowledges the support of the UK's Science and Technology Facilities Council (STFC) Ernest Rutherford Fellowship. ARM acknowledges support from Grant RYC-2016-20254 funded by MCIN/AEI/10.13039/501100011033 and by ESF Investing in your future and from MINECO under the PID2020-117252GB-I00 grant. VSD, HiPERCAM, and ULTRACAM are supported by the STFC. IP and TRM acknowledge support from the STFC, grant ST/T000406/1 and a Leverhulme Research Fellowship. JM was supported by funding from a Science and Technology Facilities Council (STFC) studentship. Based on observations collected at the European Southern Observatory under ESO programme 0106.D-0824. Based on observations made with the Gran Telescopio Canarias (GTC), installed in the Spanish Observatorio del Roque de los Muchachos of the Instituto de Astrofísica de Canarias, in the island of La Palma. This work has made use of data from the European Space Agency (ESA) mission {\it Gaia} (\url{https://www.cosmos.esa.int/gaia}), processed by the {\it Gaia} Data Processing and Analysis Consortium (DPAC, \url{https://www.cosmos.esa.int/web/gaia/dpac/consortium}). Funding for the DPAC has been provided by national institutions, in particular the institutions participating in the {\it Gaia} Multilateral Agreement. For the purpose of open access, the author has applied a Creative Commons Attribution (CC BY) licence to any Author Accepted Manuscript version arising. We thank the anonymous referee for their
helpful comments.

\section*{Data Availability}

The data underlying this article will be shared upon reasonable request to the corresponding author.



\bibliographystyle{mnras}
\bibliography{paper} 




\appendix

\section{Observations}
\label{sec:appendix_JOO}
\begin{table*}
    \centering
    \begin{tabular}{@{}lccccccc@{}}
    \hline
        Target & Date at & Start & Telescope-Instrument & Filters & Exposure & Number of & Conditions \\
               & start of run & (UT) & & & time (s) & exposures & (Transparency, seeing) \\ 
        \hline
        ZTF\,J012607.79$+$121049.1 & 2021-07-08 & 09:03:28 & NTT-ULTRACAM & $u_{s}g_{s}i_{s}$ & 7.0 & 804 & clear, $\sim$1 arcsec \\
        ZTF\,J022004.55$+$630359.6 & 2021-08-07 & 04:40:29 & GTC-HiPERCAM & $u_{s}g_{s}r_{s}i_{s}z_{s}$ & 0.7 & 3568 & clear, $\sim$0.6 arcsec \\
        ZTF\,J040627.23$+$095827.0 & 2021-02-08 & 01:10:49 & NTT-ULTRACAM & $u_{s}g_{s}i_{s}$ & 7.5 & 269 & clear, $<$1.5 arcsec \\
                          & 2021-11-08 & 02:54:14 & NTT-ULTRACAM & $u_{s}g_{s}i_{s}$ & 8.0 & 1426 & clear, $\sim$1 arcsec \\
                          & 2021-11-10 & 06:25:25 & NTT-ULTRACAM & $u_{s}g_{s}r_{s}$ & 8.0 & 712 & clear, $<$1.5 arcsec \\
                          & 2021-09-13 & 02:10:27 & GTC-HiPERCAM & $u_{s}g_{s}r_{s}i_{s}z_{s}$ & 1.2 & 3183 & clear, $<$2.0 arcsec \\
        ZTF\,J041016.82$-$083419.5 & 2022-03-08 & 00:51:41 & NTT-ULTRACAM & $u_{s}g_{s}i_{s}$ & 3.8 & 528 & clear, $<$1.8 arcsec \\
        ZTF\,J051902.06$+$092526.4 & 2021-02-06 & 00:59:07 & NTT-ULTRACAM & $u_{s}g_{s}i_{s}$ & 8.0 & 156 & clear, $<$1.5 arcsec \\
        ZTF\,J052848.24$+$215629.0 & 2021-02-08 & 01:51:58 & NTT-ULTRACAM & $u_{s}g_{s}i_{s}$ & 4.0 & 632 & clear, $<$1.4 arcsec \\
                                   & 2022-12-19 & 04:41:40 & NTT-ULTRACAM & $u_{s}g_{s}i_{s}$ & 3.7 & 1806 & clear, $<$2 arcsec \\
        ZTF\,J053708.26$-$245014.6 & 2021-02-06 & 04:42:38 & NTT-ULTRACAM & $u_{s}g_{s}i_{s}$ & 3.0 & 1386 & clear, $<$1.3 arcsec \\
        ZTF\,J061530.96$+$051041.8 & 2021-03-07 & 02:01:48 & NTT-ULTRACAM & $u_{s}g_{s}i_{s}$ & 6.8 & 774 & clear, $<$1.4 arcsec \\
        ZTF\,J061809.92$-$091904.3 & 2022-12-19 & 01:07:56 & NTT-ULTRACAM & $u_{s}g_{s}i_{s}$ & 4.5 & 690 & clear, $\sim$1.3 arcsec \\
                          & 2022-12-19 & 08:00:21 & NTT-ULTRACAM & $u_{s}g_{s}i_{s}$ & 3.4 & 474 & clear, $\sim$1.3 arcsec \\
        ZTF\,J063808.71$+$091027.4 & 2021-02-07 & 02:57:13 & NTT-ULTRACAM & $u_{s}g_{s}i_{s}$ & 7.0 & 656 & clear, $<$1.2 arcsec \\
        ZTF\,J063954.70$+$191958.0 & 2021-02-06 & 03:34:25 & NTT-ULTRACAM & $u_{s}g_{s}i_{s}$ & 3.0 & 1300 & clear, $\sim$1.3 arcsec \\
        ZTF\,J064242.41$+$131427.6 & 2021-02-07 & 01:52:06 & NTT-ULTRACAM & $u_{s}g_{s}i_{s}$ & 4.5 & 591 & clear, $\sim$1 arcsec \\
        ZTF\,J065103.70$+$145246.2 & 2021-02-07 & 04:36:41 & NTT-ULTRACAM & $u_{s}g_{s}i_{s}$ & 7.0 & 359 & clear, $\sim$1.5 arcsec \\
        ZTF\,J070458.08$-$020103.3 & 2021-02-08 & 04:45:27 & NTT-ULTRACAM & $u_{s}g_{s}i_{s}$ & 6.0 & 411 & clear, $<$1.5 arcsec \\
        ZTF\,J071759.04$+$113630.2 & 2021-02-06 & 02:39:58 & NTT-ULTRACAM & $u_{s}g_{s}i_{s}$ & 4.0 & 666 & clear, $<$1.3 arcsec \\
        ZTF\,J071843.68$-$085232.1 & 2021-03-10 & 01:54:13 & NTT-ULTRACAM & $u_{s}g_{s}i_{s}$ & 7.5 & 379 & clear, $<$1.2 arcsec \\
        ZTF\,J080441.95$-$021545.7 & 2022-03-08 & 01:43:23 & NTT-ULTRACAM & $u_{s}g_{s}i_{s}$ & 3.1 & 736 & clear, $\sim$1.7 arcsec \\
        ZTF\,J080542.98$-$143036.3 & 2022-03-08 & 03:30:59 & NTT-ULTRACAM & $u_{s}g_{s}i_{s}$ & 6.0 & 757 & clear, $\sim$1 arcsec \\
        ZTF\,J094826.35$+$253810.6 & 2021-01-25 & 05:43:23 & NTT-ULTRACAM & $u_{s}g_{s}i_{s}$ & 6.5 & 602 & clear, $\sim$1 arcsec \\
        ZTF\,J102254.00$-$080327.3 & 2021-03-07 & 05:35:30 & NTT-ULTRACAM & $u_{s}g_{s}i_{s}$ & 4.0 & 1224 & clear, $<$1.4 arcsec \\
        ZTF\,J102653.47$-$101330.3 & 2021-01-23 & 05:28:43 & NTT-ULTRACAM & $u_{s}g_{s}i_{s}$ & 5.8 & 316 & clear, $\sim$2 arcsec \\
        ZTF\,J103448.82$+$005201.9 & 2021-02-08 & 05:33:49 & NTT-ULTRACAM & $u_{s}g_{s}i_{s}$ & 5.8 & 408 & clear, $\sim$1 arcsec \\
        ZTF\,J104906.96$-$175530.7 & 2021-01-24 & 08:01:57 & NTT-ULTRACAM & $u_{s}g_{s}i_{s}$ & 4.5 & 866 & clear, $\sim$1 arcsec \\
        ZTF\,J120615.74+510046.8 & 2021-05-08 & 21:54:23 & GTC-HiPERCAM & $u_{s}g_{s}r_{s}i_{s}z_{s}$ & 5.4 & 9509 & thin cloud, $\sim$2 arcsec \\
        ZTF\,J122009.98$+$082155.0 & 2021-02-07 & 05:28:34 & NTT-ULTRACAM & $u_{s}g_{s}i_{s}$ & 7.0 & 536 & clear, $<$1.6 arcsec \\
        ZTF\,J125620.57$+$211725.8 & 2022-03-03 & 07:38:55 & NTT-ULTRACAM & $u_{s}g_{s}i_{s}$ & 2.5 & 1440 & clear, $\sim$1 arcsec \\
        ZTF\,J130228.34$-$003200.2 & 2021-02-07 & 08:39:46 & NTT-ULTRACAM & $u_{s}g_{s}i_{s}$ & 4.0 & 692 & clear, $\sim$1 arcsec \\
        ZTF\,J134151.70$-$062613.9 & 2021-01-23 & 06:19:18 & NTT-ULTRACAM & $u_{s}g_{s}i_{s}$ & 3.5 & 1216 & clear, $\sim$1.8 arcsec \\
        ZTF\,J140036.65$+$081447.4 & 2021-02-08 & 06:15:28 & NTT-ULTRACAM & $u_{s}g_{s}i_{s}$ & 10.0 & 434 & clear, $\sim$1 arcsec \\
        ZTF\,J140423.86$+$065557.7 & 2021-01-25 & 07:56:09 & NTT-ULTRACAM & $u_{s}g_{s}i_{s}$ & 6.0 & 705 & clear, $\sim$1 arcsec \\
        ZTF\,J140537.34$+$103919.0 & 2021-02-06 & 07:28:35 & NTT-ULTRACAM & $u_{s}g_{s}i_{s}$ & 10.0 & 276 & clear, $\sim$1.2 arcsec \\
        ZTF\,J140702.57$+$211559.7 & 2021-02-07 & 07:22:27 & NTT-ULTRACAM & $u_{s}g_{s}i_{s}$ & 4.5 & 468 & clear, $\sim$1.2 arcsec \\
                                   & 2022-03-05 & 07:35:10 & NTT-ULTRACAM & $u_{s}g_{s}i_{s}$ & 6.0 & 1304 & clear, $\sim$1 arcsec \\
                                   & 2022-03-27 & 04:07:31 & NTT-ULTRACAM & $u_{s}g_{s}r_{s}$ & 5.0 & 3455 & clear, $\sim$1.1 arcsec \\
        ZTF\,J145819.54$+$131326.7 & 2021-02-07 & 08:05:49 & NTT-ULTRACAM & $u_{s}g_{s}i_{s}$ & 6.5 & 284 & clear, $\sim$1 arcsec \\
        ZTF\,J162644.18$-$101854.3 & 2021-02-08 & 08:18:21 & NTT-ULTRACAM & $u_{s}g_{s}i_{s}$ & 5.0 & 660 & clear, $\sim$1.3 arcsec \\
        ZTF\,J163421.00$-$271321.7 & 2021-02-08 & 07:35:28 & NTT-ULTRACAM & $u_{s}g_{s}i_{s}$ & 5.4 & 420 & clear, $\sim$1.2 arcsec \\
        ZTF\,J164441.18$+$243428.2 & 2021-03-10 & 09:06:00 & NTT-ULTRACAM & $u_{s}g_{s}i_{s}$ & 6.0 & 321 & clear, $\sim$1.0 arcsec \\
        ZTF\,J180256.45$-$005458.3 & 2022-03-08 & 08:08:10 & NTT-ULTRACAM & $u_{s}g_{s}i_{s}$ & 4.0 & 728 & clear, $\sim$1.2 arcsec \\
        ZTF\,J182848.77$+$230838.0 & 2022-04-26 & 09:09:02 & NTT-ULTRACAM & $u_{s}g_{s}i_{s}$ & 4.0 & 532 & clear, $\sim$2 arcsec \\
        ZTF\,J192215.32$+$103838.1 & 2022-06-07 & 06:46:35 & NTT-ULTRACAM & $u_{s}g_{s}i_{s}$ & 3.9 & 803 & thin cloud, $\sim$1 arcsec \\
        ZTF\,J195456.71$+$101937.5 & 2022-04-28 & 07:19:49 & NTT-ULTRACAM & $u_{s}g_{s}i_{s}$ & 1.4 & 3174 & clear, $<$2.5 arcsec \\
        ZTF\,J214232.02$+$430929.0 & 2021-09-06 & 23:17:08 & GTC-HiPERCAM & $u_{s}g_{s}r_{s}i_{s}z_{s}$ & 2.0 & 1544 & clear, $\sim$1 arcsec \\
        ZTF\,J222007.49$+$072129.7 & 2021-09-11 & 00:04:54 & GTC-HiPERCAM & $u_{s}g_{s}r_{s}i_{s}z_{s}$ & 0.7 & 4047 & some clouds, $\sim$0.8 arcsec \\
        ZTF\,J235354.98$+$415304.4 & 2021-09-09 & 04:43:15 & GTC-HiPERCAM & $u_{s}g_{s}r_{s}i_{s}z_{s}$ & 1.0 & 2770 & clear, $\sim$0.6 arcsec \\
        \hline
    \end{tabular}
    \caption{Journal of observations.}
    \label{tab:observations}
\end{table*}

\section{Light curves}
\label{sec:light_curves}

\begin{figure*}
 \includegraphics[width=\textwidth]{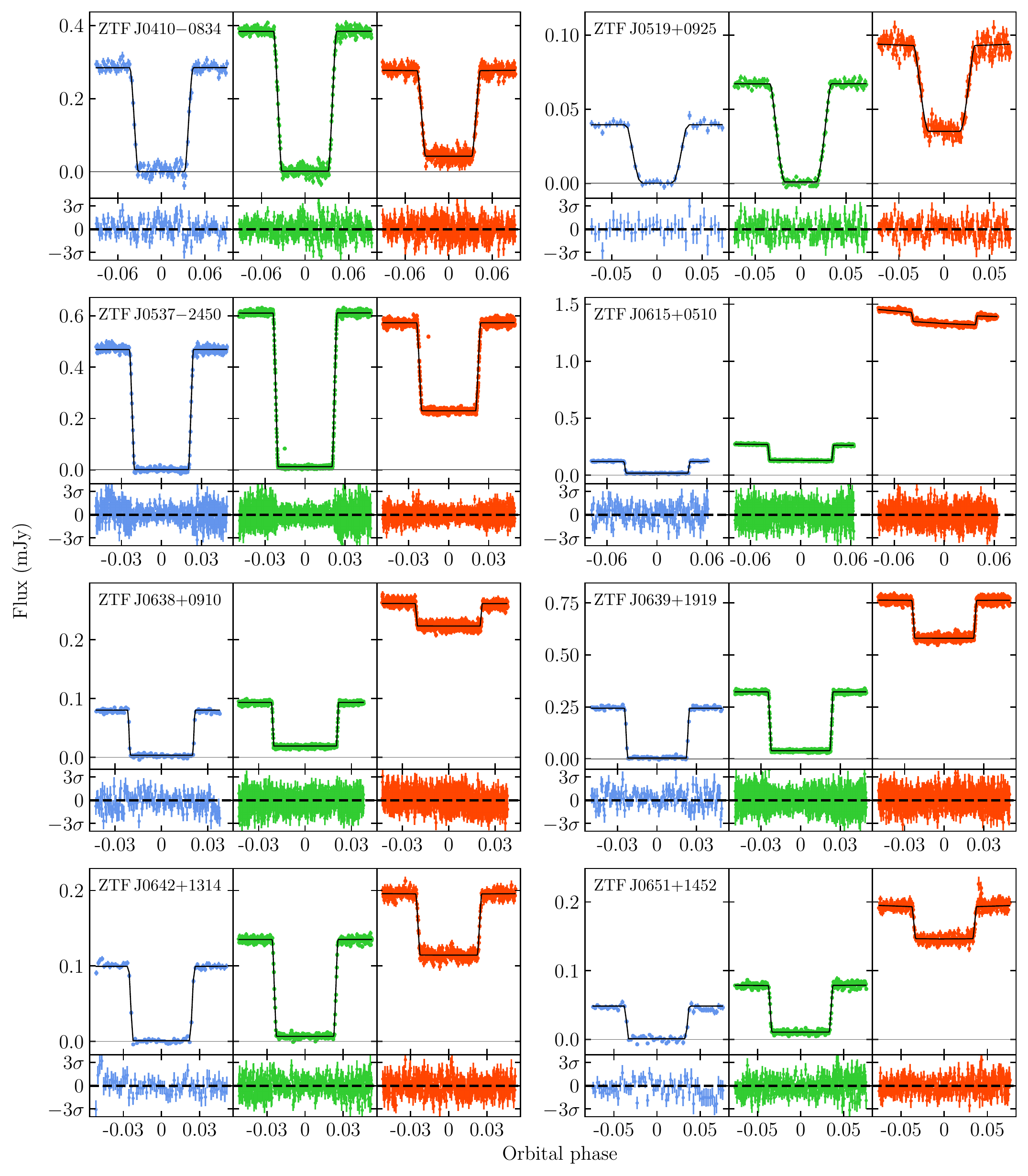}
 \caption{Best-fit light curve models (solid black lines) to the ULTRACAM $u_{s}\,g_{s}\,i_{s}$ eclipse photometry (coloured points) of the non-pulsating systems. The horizontal grey lines show a flux of zero. Residuals of the best-fit models are shown in the panels below the light curves.}
 \label{fig:light_curves_1}
\end{figure*}

\begin{figure*}
 \includegraphics[width=\textwidth]{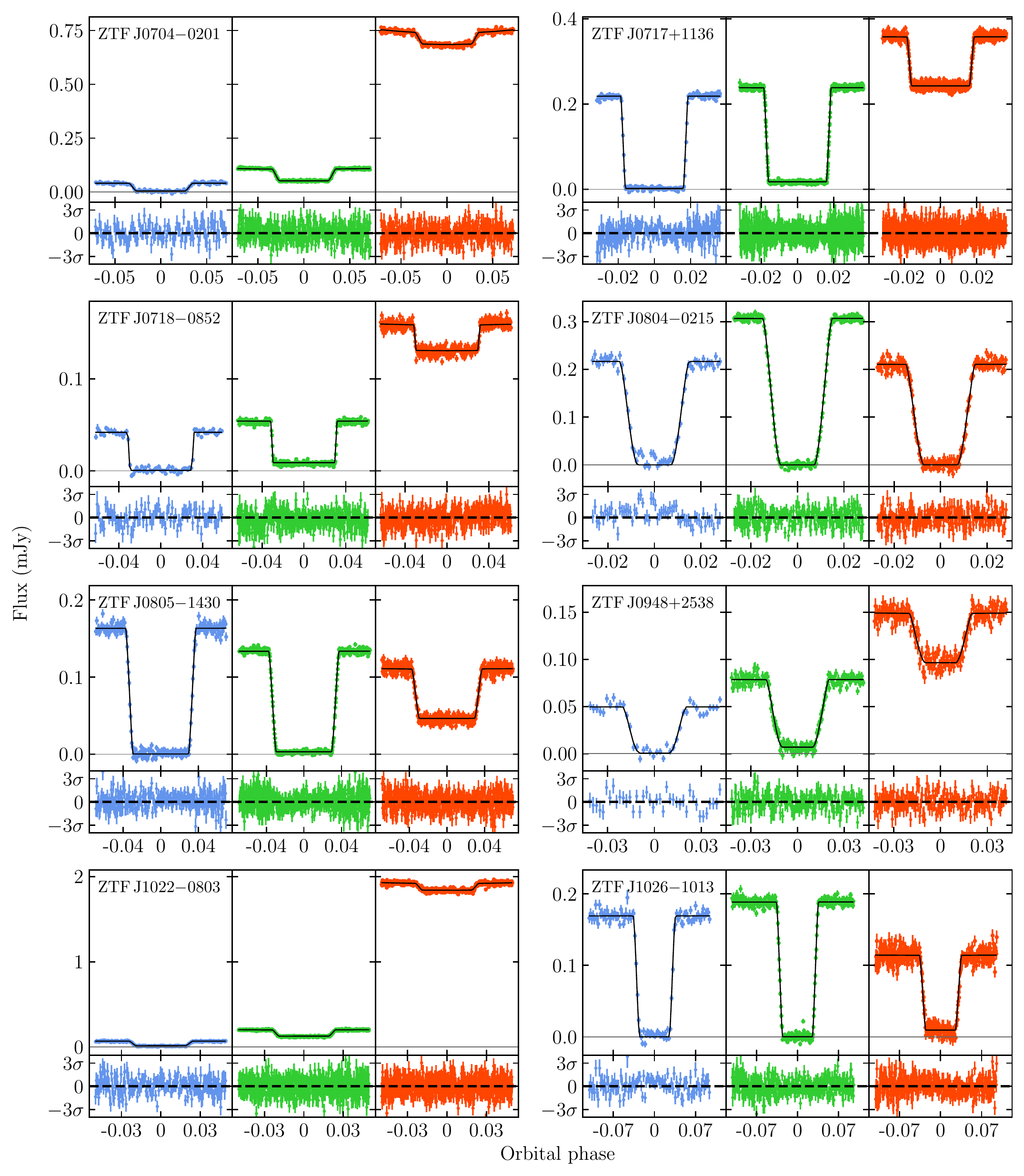}
 \caption{As in \autoref{fig:light_curves_1}.}
 \label{fig:light_curves_2}
\end{figure*}

\begin{figure*}
 \includegraphics[width=\textwidth]{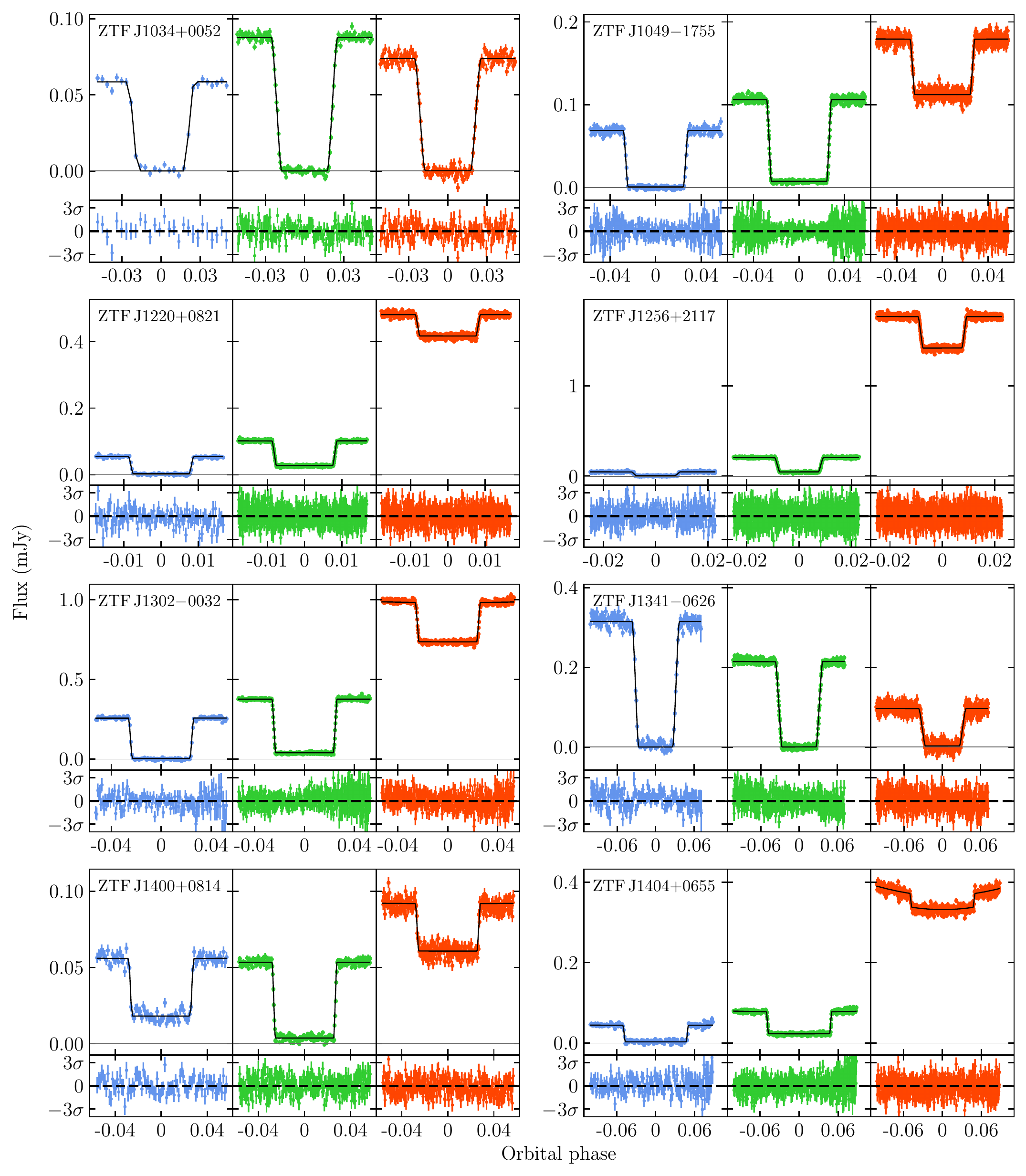}
 \caption{As in \autoref{fig:light_curves_1}.}
 \label{fig:light_curves_3}
\end{figure*}

\begin{figure*}
 \includegraphics[width=\textwidth]{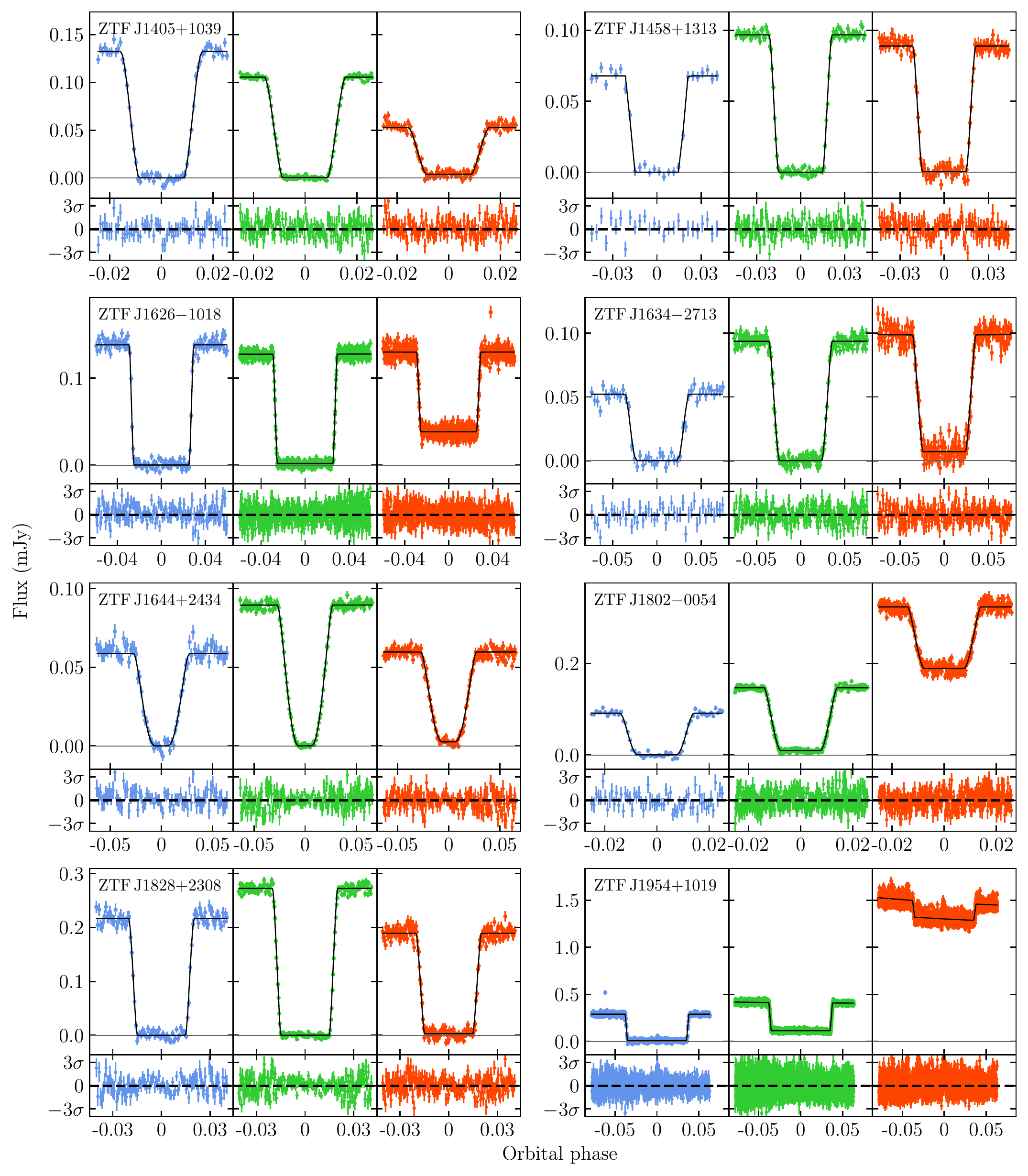}
 \caption{As in \autoref{fig:light_curves_1}.}
 \label{fig:light_curves_4}
\end{figure*}


\bsp	
\label{lastpage}
\end{document}